\documentclass[12pt]{article}
\usepackage{latexsym}
\usepackage{amssymb}
\usepackage{graphicx}
\usepackage{multibox}

\topskip \textwidth 480pt

\def\*{\ast}
\def\a{\alpha}

\def\go{\omega}

\def\ga{{{\alpha}}}

\def\D{\Delta}

\def\ie{{\it i.e.,\,\,}}

\def\o{\omega}
\def\s{\sigma}

\def\t{\tau}

\def\pa{\partial}

\def\be{\begin{equation}}
\def\ee{\end{equation}}
\def\bqn{\begin{eqnarray}}
\def\eqn{\end{eqnarray}}
\def\nn{\nonumber}

\def\gvep{\varepsilon}

\def\gb{\beta}
\def\gga{\gamma}
\def\theequation{\thesection.\arabic{equation}}

\def\f{\frac}

\newsavebox{\ver}
\newsavebox{\verp}
\newsavebox{\gorp}
\newsavebox{\toch}

\newcommand{\bee}{\begin{eqnarray}}
\newcommand{\eee}{\end{eqnarray}}

\newcommand{\FR}{{Fronsdal~}}

\newcommand\ls{\!\!\!\!\!\!\!}

\def\q{{\,,\qquad}}

\def\bY{{\bar{Y}}}
\def\bZ{{\bar{Z}}}

\date{}
\begin{document}
\begin{titlepage}
\title{
\begin{flushright}
{\small FIAN/TD/20-08}\\
~\\
~\\
\end{flushright}
{\bf Reducible higher--spin multiplets in flat and AdS spaces and
their geometric frame--like formulation}
~\\
\author{D.P.~Sorokin$^*$ and M.A.~Vasiliev$^\dagger$
~\\
~\\
{\it $^*$ Istituto Nazionale di Fisica Nucleare, Sezione di Padova,}
~\\
{\it via F. Marzolo 8, 35131 Padova, Italia}\\
~\\
{\it $^\dagger$ I.E.Tamm Department of Theoretical Physics, Lebedev
Physical Institute,}
~\\
{\it Leninsky prospect 53, 119991 Moscow, Russia} }}

\maketitle
\begin{abstract}
\noindent We consider the frame--like formulation of reducible
sets of totally symmetric bosonic and fermionic higher--spin  fields
in flat and AdS backgrounds of any dimension, that correspond to
so-called higher--spin triplets resulting from the string--inspired
BRST approach. The explicit relationship of the fields of
higher--spin triplets to the higher--spin vielbeins and connections
is found. The gauge invariant actions are constructed including, in
particular, the reducible (\emph{i.e.} triplet) higher--spin fermion
case in $AdS_D$ space.
\end{abstract}
\thispagestyle{empty}
\end{titlepage}

\tableofcontents
\section{Introduction}
This paper is an essentially extended version of the contribution to
the volume dedicated to the 60th birthday anniversary of Joseph
Buchbinder, our colleague and friend, who, among other important
subjects in his fruitful scientific carrier, made an extensive
contribution to the theory of higher--spin fields.

The minimal approach to the description of massless higher--spin
fields, developed originally by Fronsdal
\cite{fronsdal78,fronsdalAdS}
 and Fang and Fronsdal \cite{fronsdal78a,fronsdal78b}
for the generic massless fields in four dimensional Minkowski
\cite{fronsdal78,fronsdal78a} and anti--de Sitter space
\cite{fronsdalAdS,fronsdal78b}, is usually referred to as the
metric--like formalism because it is a natural generalization
\cite{WF} of the
metric formulation of the linearized gravity  (\ie massless spin
two). The construction of gauge invariant actions for single
(irreducible) massless higher--spin fields in this approach requires
 these fields to be double--traceless in the
bosonic case
\cite{fronsdal78,fronsdalAdS} or triple--gamma--traceless
\cite{fronsdal78a,fronsdal78b} in the fermionic case.

The frame-like formulation of massless higher--spin gauge fields,
that generalizes the Cartan formulation of gravity, is also
available both in Minkowski
\cite{V80,DA} and anti--de--Sitter \cite{V87,LV,Vasiliev:1987tk,5d} spaces.
In this approach, higher--spin fields are described by differential
forms that carry  irreducible representations of the fiber Lorentz
group. In the spin--two case this approach reproduces the linearized
Cartan gravity with the one-form frame field or vielbein
$e^a=dx^me_m{}^a$ carrying a vector representation (index $a$) of
the Lorentz group. For higher spins, the frame-like fields are rank
$s-1$ symmetric traceless tensors $e^{a_1\ldots
a_{s-1}}=dx^me_m{}^{a_1\ldots a_{s-1}}$. Higher-spin gauge symmetry
parameters $\xi^{a_1\ldots a_{s-1}}$ are rank $s-1$ traceless
 symmetric tensors in the both approaches.

Both metric-like and frame-like  approaches are geometric, although
in a slightly different fashion,  extending Riemann and Cartan
geometries, respectively. As in the standard case of gravity, the
frame-like approach
\cite{V80,DA,V87,LV,Vasiliev:1987tk,5d} is more general
than the metric-like
\cite{fronsdal78,fronsdalAdS,fronsdal78a,fronsdal78b,WF}. The latter
is a particular gauge of the former. Moreover, in the fermionic
case, the frame-like approach is the only one working at the
interaction level. In fact, the frame-like approach, which operates
in terms of differential forms and connections, has a greater power.
It allows, in particular via the unfolded dynamics approach
\cite{Ann} (see \cite{bciv} for more detail), to introduce
higher-spin interactions and to uncover deep geometric structures
that underly the higher--spin theory and are likely to deviate from
the standard concepts of Riemann geometry in the strong field
regime. (For more detail on the higher--spin theory see recent
reviews
\cite{bciv,MV0401,DS04,bcs,Sagnotti:2005ns,Fotopoulos:2008ka} and
references therein).

If, instead of a single spin, a set of different spins is
considered, their dynamics can be described in terms of tensor
fields that are less constrained than in the single higher--spin
case or even unconstrained. An example of such a system is provided
by the so called triplet systems of massless higher--spin fields
which naturally appear in the process of truncation of the open
string spectrum in the tensionless limit
\cite{Ouvry:1986dv,Bengtsson:1986ys,ht,Pashnev:1989gm,Francia:2002pt,Sagnotti:2003qa,Barnich:2005ga}
(see also \cite{Buchbinder:2006eq} for further developments and
e.g.
\cite{Gross:1988ue,Sundborg:2000wp,Lindstrom:2003mg,Bonelli:2003kh}
for other aspects of the tensionless string limit and higher spin
theory). So one can regard the triplets as fields which manifest
their origin from massive higher--spin fields of the tensionful
string.

The geometrical nature of triplet fields, \emph{i.e.} their relation
to higher--spin counterparts of metric (or vielbein) and connection,
has not been clarified yet. Moreover, neither equations of motion
nor the action for fermionic triplets in AdS space have been
constructed. This hinders the study of the relation of the fermionic
triplets to string states in AdS backgrounds and corresponding
applications.

A purpose of this paper is to reconsider these problems using the
frame--like approach. Upon establishing the geometrical meaning of
the triplet fields and finding their proper gauge transformations
both in Minkowski and in $AdS_D$, we construct the Lagrangian
description of the bosonic and fermionic triplets in flat and AdS
backgrounds. We present two descriptions of AdS triplet systems. The
formulation of Section \ref{flmAdS}, which uses $O(1,D-1)$ Lorentz
tensors as in
\cite{V80} for irreducible fields, is relatively simple but is only
implicitly gauge invariant while another one considered in Section
\ref{ADSC}, as in
\cite{5d} for irreducible fields, is manifestly gauge and $AdS_D$
invariant (\emph{i.e.} invariant under the $AdS_D$ isometry group
$O(2,D-1)$) but requires a somewhat more involved action.  Note that
the relaxed systems of fields that contain the triplet systems along
with
 the so-called partially
massless fields \cite{pm} in the frame--like formalism \cite{SV}
were considered recently by Alkalaev in \cite{alkal} within the
$AdS_D$ covariant frame-like approach. We extend these results by
constructing actions and formulating conditions that sort out the
quantum-mechanically inconsistent (nonunitary in anti - de Sitter,
or tachionic in de Sitter space)
 partially massless fields.

As we have mentioned, to describe the dynamics of a field of a
single higher--spin in the frame--like formulation one should impose
traceless conditions on higher--spin vielbeins and connections
\cite{V80,DA,V87,LV,Vasiliev:1987tk,5d,MV0401,bciv}. We shall show that these conditions
can be relaxed in such a way that the higher--spin vielbein becomes
unrestricted, while the higher--spin connections are subject to
weaker traceless constraints. In the integer--spin case  the weaker
conditions result in a Lagrangian system that describes the set of
fields of spins $s$, $s-2, s-4, \ldots , 3$ or 2, depending on
whether $s$ is odd or even. The system formulated in terms of
frame-like differential forms does not describe lower--spin massless
fields with spins $s\leq 1$. These can be included by adding to the
action their kinetic terms formulated in terms of the so--called
Weyl zero--forms as discussed in \cite{SHV} for the scalar case and
in \cite{fvnp} for the spin one case. Analogously, in the
half--integer spin case, the system under consideration describes
the set of fields of spins $s, s-1, s-2,\ldots, 3/2$.  The
description a massless spin 1/2 field also needs inclusion of
zero-forms. To simplify consideration we discard the analysis of the
lower--spin fields in this paper. Let us stress that we consider the
reducible fermionic systems both in flat and in $AdS$ spaces, that
is important for the analysis of a relationship of higher--spin
theories with superstrings.

The higher--spin triplets described in the metric--like formulation
have the same physical state contents (modulo the lowest spin states
0, 1 and 1/2), and indeed we find the correspondence between the
triplet fields and components of the higher--spin vielbein and
connection of the higher--spin system with relaxed trace
constraints. A transparent geometrical structure of the frame--like
formulation allows us to construct relatively simple actions for the
bosonic and, in particular, fermionic triplet fields both in flat
and AdS backgrounds which should be useful for their applications,
\emph{e.g.} for studying interactions of triplets.

The trace constraints on the dynamical fields and gauge symmetry
parameters, which single out the irreducible higher--spin fields,
are algebraic (free of derivatives of fields) and therefore
harmless. If desired, they can be easily removed by introducing
Lagrange multipliers along with Stueckelberg fields and symmetries
to make the higher--spin gauge fields and parameters traceful.
Several versions of the unconstrained Lagrangian formulation of
higher--spin fields have been proposed in the literature using
different approaches (see
\emph{e.g.}
\cite{Pashnev:1997rm,Burdik:2001hj,Buchbinder:2001bs,Francia:2002aa,
Francia:2002pt,
Francia:2005bu,Buchbinder:2006ge,Buchbinder:2007ak,Buchbinder:2007vq,Moshin:2007jt,Francia:2007ee,
Barnich:2004cr}). We shall also show how imposing constraints on the
(gamma)--trace of the higher--spin vielbeins to be pure gauge
reduces the higher--spin systems under consideration to the
frame--like versions of the unconstrained formulations of single
higher--spin fields considered in
\cite{Buchbinder:2007ak,Francia:2007ee}.

The triplet systems of higher-spin fields resulting from string
theory in the tensionless limit by construction carry the sets of
states that are appropriate for the description of massive
higher--spin fields. In other words, the study of these sets may
shed some light on a mechanism of higher--spin symmetry breaking
resulting in the generation of mass in Higher--Spin Theory, which is
the necessary step in establishing a relationship of the
Higher--Spin Gauge Theory with String Theory. In particular, the
results of this work are expected to help to obtain frame--like
versions of the string--inspired BRST formulation of massless and
massive higher--spin fields (see \emph{e.g.}
\cite{Fotopoulos:2008ka} for a recent review and references) as well
as of the gauge invariant (Stueckelberg) description of the massive
fields analogous to the metric--like approach by Zinoviev
\cite{zin,Metsaev:2006zy} both in the bosonic and fermionic cases.
Note that the BRST version of the frame-like unfolded formulation of
massless higher--spin fields was considered in
\cite{Barnich:2004cr} for the Minkowski case and in
\cite{Barnich:2006pc} for the AdS case.

A closely related problem for the future study is to figure out what
might be an algebraic structure (higher--spin symmetry) that
underlies the relaxed higher--spin multiplets considered in this
paper. In this respect, an encouraging result is that the relaxed
higher--spin systems under consideration can be singled out from the
appropriate oscillator algebras by imposing natural conditions which
are invariant under the $AdS_D$ symmetry algebra $o(2,D-1)$. These
conditions pick up only those representations of $o(2,D-1)$ that
correspond to standard (unitary) systems of massless fields, sorting
out the non--unitary systems that describe the partially massless
fields \cite{pm} in the frame--like formalism \cite{SV}.

The results of this paper can be useful for the further study of
higher--spin triplets and their generalization to mixed--symmetry
fields in particular in AdS backgrounds (on various aspects of
mixed--symmetry fields see
\emph{e.g.} \cite{{Ouvry:1986dv},{Labastida:1987kw},{Metsaev:1997nj},
{Brink:2000ag},{Burdik:2001hj},{Bekaert:2002dt},{de
Medeiros:2002ge},{Zinoviev:2002ye},{Buchbinder:2007vq},Moshin:2007jt}
and references therein), in various contexts of higher--spin theory
and its relation to string theory.

For the structure of the paper see the Table of Contents.

\section{Frame--like action for bosonic higher--spin fields in flat space--time} \label{flm}

In the frame--like formulation (we refer the reader to
\cite{V80,V87,MV0401,bciv} for details) a massless symmetric field of an integer spin $s$
in flat space--time of dimension $D$ is described by the
higher--spin vielbein one--form\footnote{In flat space--time we
shall not distinguish between the world and tangent--space indices.
Both kinds of indices will be denoted by  lower case Latin letters.
World indices will be separated  from the tangent--space ones by
`;'.}
\begin{equation}\label{viel}
e^{n_1 \ldots n_{s-1}}=dx^m\,e_{m;}{}^{n_1 \ldots n_{s-1}},
\end{equation}
by the one--form connection
\begin{equation}\label{connect}
\omega^{n_1 \ldots n_{s-1},p}=dx^m\,\omega_{m;}{}^{n_1 \ldots n_{s-1},p},
\end{equation}
and by so-called extra fields that do not contribute to the
 free action and field equations but control higher--spin gauge symmetries
 and play a role at the interaction
 level \cite{fvnp,5d}. In (\ref{viel}) and (\ref{connect}) the
indices $n_1 \ldots n_{s-1}$ are symmetrized, and $\omega^{n_1
\ldots n_{s-1},p}$ has the symmetry properties of the Young tableau
$Y(s-1,1)$ \footnote{ Sets of symmetric tangent--space indices are
separated by comma. $Y(s-1,1)$ means that the Young tableau has
$s-1$ cells in the first row and one cell in the second row.},
\emph{i.e.} its  totally symmetric part in tangent--space indices vanishes
\begin{equation}\label{ts}
\omega^{(n_1 \ldots n_{s-1},p)}:={1\over s}\,(\omega^{n_1 \ldots n_{s-1},p}+\omega^{p \ldots n_{s-1},n_1}
+ s-2~{\rm terms})=0\,.
\end{equation}
The brackets $()$ and $[]$ denote, respectively,  the
symmetrization and anti--symmetrization of indices with the unit
weight.

The connection $\omega^{n_1 \ldots n_{s-1},p}$ is an auxiliary field
provided that, like in the case of the Einstein gravity, we impose
the zero torsion condition
\begin{equation}\label{zt}
T^{n_1 \ldots n_{s-1}}\equiv d\,e^{n_1 \ldots
n_{s-1}}-{(s-1)}\,dx^q\,\omega^{n_1 \ldots
n_{s-1},p}\,\eta_{pq}=0\,.
\end{equation}
The dynamical degrees of freedom of a massless higher--spin field
are contained in the higher--spin vielbein (\ref{viel}) which also
describes pure gauge degrees of freedom because of the presence of
the higher--spin gauge symmetries. In particular,  the torsion
(\ref{zt}) is invariant under the following gauge transformations of
the vielbein and connection
\begin{equation}\label{gs1}
\delta \,e^{n_1 \ldots n_{s-1}}=d\xi^{n_1 \ldots n_{s-1}}-(s-1)\,dx^q\,\xi^{n_1 \ldots
n_{s-1},p}\,\eta_{pq}\,,
\end{equation}
\begin{equation}\label{gs2}
\delta\, \omega^{n_1 \ldots
n_{s-1},p}= d\xi^{n_1 \ldots n_{s-1},p}-(s-2)\,dx^q\,\xi^{n_1 \ldots
n_{s-1},pr}\,\,\eta_{rq}\,.
\end{equation}
The gauge parameters $\xi^{n_1 \ldots n_{s-1}}$, $\xi^{n_1 \ldots
n_{s-1},p}$ and  $\xi^{n_1 \ldots n_{s-1},p_1p_2}$ are symmetric in
each group of indices $n$ and $p$.  In addition, $\xi^{n_1 \ldots
n_{s-1},p}$ and  $\xi^{n_1 \ldots n_{s-1},p_1p_2}$ have the symmetry
properties of the Young tableaux $Y(s-1,1)$ and $Y(s-1,2)$,
respectively, which means, like in (\ref{ts}), that the
symmetrization of any $s$ indices gives zero. Note that the gauge
symmetry parameter $\xi^{n_1 \ldots n_{s-1},p_1p_2}$ is associated
with the first extra field connection $\go^{n_1 \ldots
n_{s-1},p_1p_2}$ that has analogous symmetry properties in the
tangent indices.

Note that so far we have not imposed traceless conditions either on
the higher--spin vielbein and connection or on the gauge parameters.

We would like to derive the zero torsion condition (\ref{zt}) from
an action together with dynamical field equations on the physical
components of $e^{n_1 \ldots n_{s-1}}$. We construct such an action
by analogy with the frame formulation of the action for (linearized)
gravity.

The free higher--spin action has the following  simple first--order
 form
%
\be \label{fract} S= \int_{M^D}\,dx^{a_1} \ldots
dx^{{a_{D-3}}}\,\varepsilon_{a_1 \ldots a_{D-3}pqr}\,(d\,e^{n_1
\ldots n_{s-2} p} -{{s-1}\over 2}\, dx_m \,\o^{n_1 \ldots n_{s-2} p,\, m})\,
\o_{n_1 \ldots n_{s-2}}{}^{q,\, r}\,. \ee
This action is a straightforward generalization of the $4d$ action
of \cite{V80}. It has the important property that its part  bilinear
in the higher--spin connection $\o^{n_1 \ldots n_{s-1},\, m}$ is
symmetric under the exchange of the product factors,\emph{ i.e.}
\be \label{exch} \hspace{-30pt} dx^{a_1} \ldots
dx^{{a_{D-3}}} dx_m
\,\varepsilon_{a_1 \ldots a_{D-3}pqr}\,
 (\o_1{}^{n_1 \ldots n_{s-2} p,\, m}\,
\o_2{}_{n_1 \ldots n_{s-2}}{}^{q,\, r} -
\o_2{}^{n_1 \ldots n_{s-2} p,\, m}\,
\o_1{}_{n_1 \ldots n_{s-2}}{}^{q,\, r} )=0\,,
\ee
provided that $\omega$ is Young projected as in eq.~(\ref{ts}) and,
in addition, is subject to the trace constraint
\be \eta_{n_{1}m}\,\o^{n_1 \ldots n_{s-1},\, m}= 0 \,.  \label{trnn}
\ee
Indeed, this can be easily seen by using the identity
\be
\label{idep} \varepsilon_{a_1\cdots\,a_{D-3}bcd}\,e^{a_1} \ldots
e^{a_{D-3}}\, e^f =
{3\over{D-2}}\,\delta^f_{[b}\,\varepsilon_{cd]\,a_1\cdots\,a_{D-2}}\,e^{a_1}
\ldots e^{a_{D-2}}\,,
\ee
where in the case under consideration the background vielbein $e^a$
is flat, \emph{i.e.} $e^a=dx^a$ in  Cartesian coordinates. The
identity (\ref{idep}) is a particular case of the generic identity
\be\label{idep1}
\varepsilon_{a_1\cdots\,a_{D-p}b_1 \cdots b_p}\,e^{a_1} \ldots
e^{a_{D-p}}\, e^f =
{{(-1)^{(p-1)(D-p+1)}\,p}\over{D-p+1}}\,\delta^f_{[b_1}\,\varepsilon_{b_2\cdots
b_p]\,a_1\cdots\,a_{D-p+1}}\,e^{a_1}
\ldots e^{a_{D-p+1}}\,,
\ee
which (for different $p$) have been used when checking the gauge
invariance of the actions considered in this paper.

Thus, the part of (\ref{fract}) bilinear in $\omega$ consists of
three terms, that contain $\eta_{mp}$, $\eta_{mq}$ and $\eta_{mr}$,
respectively. The first term vanishes by the trace condition
(\ref{trnn}), while the other two are symmetric, either manifestly
or on account of the Young projection property (\ref{ts}).

Note that a consequence of (\ref{trnn}) is that the trace of
$\omega$ in the first group of symmetrized indices $n$ has again
definite $Y(s-3,1)$  Young--symmetry  properties
\be \eta_{n_{1}n_2}\,\o^{n_1 n_2(n_3 \ldots n_{s-1},\, m)}= 0 \,.  \label{try}
\ee
Let us stress that the trace $\eta_{n_{1}n_2}\,\o^{n_1 n_2n_3
\ldots n_{s-1},\, m}$ is \emph{non--zero}. Therefore, the condition
(\ref{trnn}) is weaker than the conventional trace constraint of the
frame--like formulation of a single higher--spin field which
corresponds to Fronsdal theory \cite{fronsdal78} and requires all
traces in tangent indices to be zero. We shall call eq. (\ref{trnn})
the relaxed traceless condition.

Note that the vielbein $e^{n_1 \ldots n_{s-1}}$ remains traceful.
Here we should point out, however, that in the case of the odd
integer spins $s=2k+1$ the fully trace part of $e^{n_1 \ldots
n_{s-1}}$,
\emph{i.e.} $e^{n_1 \ldots n_{2k}}\eta_{n_1n_2}\cdots
\eta_{n_{2k-1}n_{2k}}$, does not contribute to the action (\ref{fract})
because of its differential form structure. Technically, the reason
for this is that the one--form associated with the spin--1 field
does not have external (tangent--space) indices required for the
construction of the action as an integral of a differential form.
Thus, the action (\ref{fract}) does not describe fields of spin one.

In the case of the even integer spins $s=2k$, the total trace
component of the higher--spin vielbein $e_{n;}^{n_1 \ldots
n_{2k+1}}\delta^{n}_{n_1}\eta_{n_2n_3}\cdots \eta_{n_{2k}n_{2k+1}}$
describes the conformal mode of the spin two field, which is pure
gauge in view of the gauge transformations (\ref{gs1}). Hence, the
action (\ref{fract}) does not describe scalar fields either.

To include the spin 0 and spin 1 fields into the system one should
add to the action (\ref{fract}) the corresponding Klein--Gordon and
Maxwell terms. This can be achieved by adding the spin--one and
spin--zero kinetic terms formulated in terms of the so-called Weyl
zero-forms as discussed in
\cite{SHV} for the scalar case and in \cite{fvnp} for the spin--one case.

By virtue of (\ref{exch}) and (\ref{trnn}),  the general local
variation of the action (\ref{fract}) can be presented in the
following two forms, which are equivalent up to total derivatives,
\begin{eqnarray}\label{deltaS}
 \delta{S}&=&
 \int_{M^D}\,dx^{a_1} \ldots
dx^{{a_{D-3}}}\,{\varepsilon_{a_1 \cdots a_{D-3}}}^{pqr}\, {\delta
T}_{n_1 \cdots n_{s-2} p}\, \o^{n_1 \cdots n_{s-2}}{}_{q,\, r} \\
&=& \hspace{-5pt}
 \int_{M^D}\!\! dx^{a_1} \ldots
dx^{{a_{D-3}}}\,{\varepsilon_{a_1 \cdots a_{D-3}}}^{pqr}\, ({T}_{n_1
\cdots n_{s-2} p}\, \delta\o^{n_1 \cdots n_{s-2}}{}_{q,\, r} -
\delta{e}_{n_1 \cdots n_{s-2} p}\, d \o^{n_1 \cdots n_{s-2}}{}_{q,\,
r})\, ,\nonumber
\end{eqnarray}
where the torsion $T^{n_1 \cdots n_{s-1}}$ is defined in the left
hand side of (\ref{zt}).

{}The first form of the variation is convenient for the
identification of the gauge symmetry of the action. We notice that
whereas the torsion $T$ is invariant under arbitrary unrestricted
gauge transformations, the relaxed traceless condition (\ref{trnn})
requires gauge symmetry parameters to obey the relaxed trace
constraints as well
\begin{equation}\label{trmn}
 \eta_{n_{1}m}\,\xi^{n_1 \ldots n_{s-1},\,
m}=0\, , \quad \eta_{n_{1}m}\,\xi^{n_1 \ldots n_{s-1},\, ml}=0\,.
\end{equation}

The second line of (\ref{deltaS}) yields the field equations for
$\o$ and $e$, which follow from
\be\label{eeq1}
\delta_{[b}^{m}\delta_{c}^{n}\delta_{d]}^{r} \, T_{mn;n_1 \cdots
n_{s-2}}{}^{b}\ \delta\, \o_{r;}{}^{n_1 \cdots n_{s-2} c,\, d} = 0
\,,
\ee
\be\label{eeq}
\delta_{[b}^{m}\delta_{c}^{n}\delta_{d]}^{r}
\delta{e}_{m;n_1 \cdots n_{s-2}}{}^{b}\, \partial_n \o_{r;}{}^{n_1 \cdots n_{s-2}}{}^{c,\,
d}=0\,.
\ee
%

%
%
%
%



As we explain in Appendix, the equation (\ref{eeq1}) is equivalent
to the zero--torsion condition (\ref{zt}), modulo its full trace in
the tangent space indices in the case of spin $s=2k+1$. There is no
condition on the full trace of the torsion, since, as we have
explained above, the full trace of the higher--spin vielbein does
not contribute to the action.

The zero torsion condition
\begin{equation}\label{ofc11}
(s-1)\o_{[n;}^{~n_1 \cdots
n_{s-1},b}\eta_{m]b}=\partial_{[m}\,e_{n];}^{~n_1\cdots
n_{s-1}}\,\quad 
\end{equation}
expresses the higher--spin connection in terms of the first
derivatives of the higher--spin vielbein up to the Stueckelberg
gauge transformations (\ref{gs2}).
In the case of odd $s=2k+1$, $e_{m;}^{~n_1\cdots n_{s-1}}$  in
(\ref{ofc11}) stands for the part of the vielbein whose total trace
is zero $e_{m;}^{~n_1\cdots n_{2k}}\,\eta_{a_1a_2}\cdots
\eta_{a_{2k-1}a_{2k}}=0$.
 In view of the relation (\ref{ofc11}) the equations which follow
from eq. (\ref{eeq}), namely
\begin{equation}\label{dem}
\hspace{-25pt}
\delta^m_{(b}\partial^c\omega^{d;}{}_{n_1\cdots
n_{s-2})[c,d]}+\partial_d\,\omega_{(b;n_1\cdots
n_{s-2})}{}^{[m,d]}+\partial_{(b}\,\omega^{d;}{}_{n_1\cdots
n_{s-2})[d,}{}^{m]}=0\,
\end{equation}
 are the dynamical (second--order) equations of motion of the higher--spin
vielbein field.

Let us analyze the field content of the model.

\subsection{Fronsdal case}\label{fronsdal}
Let us  first consider the case of an irreducible massless field.
Following \cite{V80} we impose on the higher--spin vielbein and
connection the \emph{strongest} trace constraints
\be \eta_{n_{1}n_2}\,\tilde e^{n_1 \ldots n_{s-1}}=0
\qquad \eta_{n_{1}n_2}\,\tilde\o^{n_1 \ldots n_{s-1},\, m}= 0 \,,  \label{trless}
\ee
where we use  $\tilde e$ and $\tilde\omega$ for the traceless fields
to distinguish them from the relaxed $e$ and $\omega$. Note that the
condition (\ref{trnn}) follows from (\ref{ts}) and (\ref{trless}),
but not vice versa.

The parameters of the gauge transformations (\ref{gs1}) and
(\ref{gs2}) of the traceless $\tilde e$ and $\tilde\omega$ are also
traceless
\be \eta_{n_{1}n_2}\,\tilde \xi^{n_1 \ldots n_{s-1}}=0,
\qquad \eta_{n_{1}n_2}\,\tilde\xi^{n_1 \ldots n_{s-1},\, m}= 0 \,, \quad
\qquad \eta_{n_{1}n_2}\,\tilde\xi^{n_1 \ldots n_{s-1},\, mp}= 0.
 \label{trless1}
\ee

Using the gauge transformations (\ref{gs1}) and (\ref{gs2}) with the
parameters $\tilde\xi$ one can gauge fix to zero the respective
``antisymmetric" parts of the components of the vielbein $\tilde e$
and of the connection $\tilde\omega$. Then, taking into account the
zero torsion condition (\ref{ofc11}), we see that $\omega$ is the
auxiliary field and all the physical degrees of freedom are
contained in the symmetric part of the vielbein
 \be
\label{frf} s\,\tilde e_{({n}_s;\, n_1 \cdots n_{s-1} )}:=
\tilde\phi_{n_1 \cdots n_s} \,,
\ee
which is double traceless because the vielbein $\tilde
e_{{n}_s;}{}^{n_1
\cdots n_{s-1} }$ is traceless in the indices $n_1\cdots n_{s-1}$.

 The remaining local symmetry is then just that of the
 Fronsdal metric--like formulation of a single
 symmetric bosonic higher--spin field in flat space--time \cite{fronsdal78}
 with the completely
symmetric traceless parameter $\tilde \xi_{n_1 \cdots n_{s-1}}$.

If we now substitute  the connection $\tilde \o$ with its expression
(\ref{ofc11}) in terms of the symmetric and {\it double traceless}
field (\ref{frf}) into the action (\ref{fract}), the resulting
action will be quadratic in the derivatives of $\tilde\phi_{n_1
\cdots n_s}$ and will be invariant under the local transformations
\begin{equation}\label{gsf}
\delta \,\tilde\phi_{n_1 \ldots n_{s}}=s\,\partial_{(n_1}\tilde \xi_{n_2 \ldots
n_{s-1})}\,
\end{equation}
with the traceless gauge parameters $\tilde\xi_{n_1
\cdots n_{s-1}}$.
In  \cite{curt} it was shown that, up to a normalization, any such
action is equivalent to the Fronsdal action  for a spin $s$ massless
gauge field
\begin{equation}\label{ba}
S=\int d^D x \left({1\over 2} \tilde \phi^{m_1\cdots
m_{s}}\,{\mathcal F}_{m_1\cdots m_{s}}-{1\over 8}\,s(s-1)\,\tilde
\phi_n^{~nm_3\cdots m_{s}}\,{\mathcal F}^p_{~pm_3\cdots m_{s}}
\right)
\end{equation}
where
\begin{equation}\label{be}
{\mathcal F}_{m_1\cdots
m_{s}}(x)\equiv\partial^2\,\tilde\phi_{m_1\cdots m_{s}}-s\,
\partial_{(m_1}\partial^n\,\tilde\phi_{m_2\cdots
m_{s})n}
+{{s(s-1)}\over 2}\,
\partial_{(m_1}\partial_{m_2}\,\tilde\phi^n_{~m_3\cdots m_{s}
)n}
\end{equation}
is the so-called  Fronsdal operator.

\subsection{Triplet case}\label{tripletcase1}
 Let us now analyze the field content of the model described by the
action (\ref{fract}) with the traceful higher--spin vielbein and with
the higher--spin connection subject to the relaxed traceless
condition (\ref{trnn}). By representing the vielbein as a sum of
traceless (lower rank) symmetric tensors, one can see that the
action (\ref{fract}) is actually the sum of the actions for the
\emph{traceless} vielbeins $\tilde e^{a_1\cdots a_{t-1}}$ and
connections $\tilde \omega^{a_1\cdots a_{t-1},b}$ with $t$ taking
even or odd values ($t=2,4,\cdots, s$ or $t=3,5,\cdots, s$)
depending on whether $s$ is even or odd,
\begin{eqnarray}\label{fracts}
&  S=
\sum_{k=1}^{[{s\over 2}]}\,\alpha(t,D)\,\int_{M^D}\,dx^{a_1} \ldots
dx^{{a_{D-3}}}\,\varepsilon_{a_1 \ldots a_{D-3}pqr}\,(d\,\tilde
e^{n_1
\ldots n_{t-2} p}\hspace{100pt}\\
& \hspace{200pt} -{{t-1}\over 2}\, dx_m \,\tilde \o^{n_1
\ldots n_{t-2} p,\, m})\,
\tilde\o_{n_1 \ldots n_{t-2}}{}^{q,\, r}\,, \nonumber
\end{eqnarray}
where $t=2k$ or $t=2k+1$,  $[{s\over 2}]$ denotes the integral part
of $s\over 2$ when $s$ is odd, $\alpha(t,D)$ are constants which
depend on space--time dimension $D$ and the rank $t$ (spin) of the
tensor fields. Thus, the sum in (\ref{fract}) is taken over even
$t=2,4,\cdots,s-2, s$ or odd $t=3,5,\cdots, s-2,s$ depending whether
$s$ is even or odd. Each of the terms of the sum (\ref{fracts}) with
a given $t$ is gauge invariant under the transformations analogous
to eqs. (\ref{gs1}), (\ref{gs2}) but with the traceless parameters
(\ref{trless1}).

As explained in Subsection \ref{fronsdal}, for a given $t$ each term
of (\ref{fracts}) describes a single free massless field of spin
$t$. Thus the action (\ref{fracts}), and hence (\ref{fract}),
describes the family of massless fields of even integer spins
$t=2,4,\cdots,s-2, s$ and of odd integer spins $t=3,5,\cdots,
s-2,s$.

These field contents are similar to the field contents of the
higher--spin triplets
\cite{Bengtsson:1986ys,ht,Francia:2002pt,Sagnotti:2003qa} (except
for the presence in the latter of the fields of the lowest spins 0
and 1). Let us now establish the relationship between the triplet
fields and the components of the higher--spin vielbein  $e$ and
connection $\omega$, thus clarifying the geometrical meaning of the
former.

Recall that the higher--spin triplet is described by the following
three symmetric tracefull tensor fields of rank $s$, $s-1$ and $s-2$
$$
\Phi_{n_1\cdots n_s}\,,\qquad C_{n_1\cdots n_{s-1}}\,, \qquad D_{n_1\cdots
n_{s-2}}\,.
$$
On the mass shell these fields satisfy the following equations
\begin{equation}\label{triplet1}
C_{n_1\cdots n_{s-1}}=\partial_m\Phi^m{}_{n_1\cdots
n_{s-1}}-(s-1)\,\partial_{(n_{s-1}}\,D_{n_1\cdots n_{s-2})}\,,
\end{equation}
\begin{equation}\label{triplet2}
\Box\,\Phi_{n_1\cdots n_s}=s\,\partial_{(n_{s}}\,C_{n_1\cdots
n_{s-1})}\,,\qquad \Box:=\partial_m\partial^m,
\end{equation}
\begin{equation}\label{triplet3}
\Box\,D_{n_1\cdots n_{s-2}}=\partial_m\,C^m{}_{n_1\cdots
n_{s-2}}\,.
\end{equation}
Eqs. (\ref{triplet1})--(\ref{triplet3}) are invariant under the
gauge transformations
\begin{equation}\label{gauge1}
\delta \Phi_{n_1\cdots n_s}=s\,\partial_{(n_{s}}\,\xi_{n_1\cdots
n_{s-1})}
\end{equation}
\begin{equation}\label{gauge2}
\delta C_{n_1\cdots n_{s-1}}=\Box\,\xi_{n_1\cdots n_{s-1}}
\end{equation}
\begin{equation}\label{gauge3}
\delta D_{n_1\cdots
n_{s-2}}=\partial_m\,\xi^m{}_{n_1\cdots n_{s-2}}\,
\end{equation}
where the unconstrained parameter $\xi_{n_1\cdots n_{s-1}}$ is
completely symmetric.

Let us now compare the gauge transformations
(\ref{gauge1})--(\ref{gauge3}) with the gauge transformations
(\ref{gs1}) and (\ref{gs2}) of the higher--spin vielbein and
connection. Using the transformation (\ref{gs1}) with the parameter
$\xi^{n_1\cdots n_{s-1},p}$ one can gauge away the part of
 the vielbein
$e_p{}^{n_1\cdots n_{s-1}}$ that corresponds to the hook Young
tableau of $\xi^{n_1\cdots n_{s-1},p}$, \emph{i.e.} the part that
satisfies the `antisymmetry' condition $e_{(p;\,n_1\cdots
n_{s-1})}=0$ and is subject to the relaxed trace constraint similar
to (\ref{trmn}). Upon imposing this gauge fixing condition the
vielbein splits into two completely symmetric tensors of rank $s$
and $s-2$
\begin{equation}\label{ephi}
e_{(n_s;\,n_1\cdots n_{s-1})}\qquad {\rm and}\qquad {{s-2}\over
s}\,\eta^{n_{s-1}p}\,(e_{p;\,n_1\cdots
n_{s-2}n_{s-1}}-e_{(n_1;\,n_2\cdots n_{s-2})n_{s-1}p})\,.
\end{equation}
Under the gauge symmetry (\ref{gs1}), (\ref{trmn})
$e_{(n_s;\,n_1\cdots n_{s-1})}$ and $e_{p;\,n_1\cdots
n_{s-2}n_{s-1}}\eta^{n_{s-1}p}$ transform in the following way
\begin{equation}\label{ephigauge}
\delta e_{(n_s;\,n_1\cdots n_{s-1})}=\partial_{(n_{s}}\,\xi_{n_1\cdots
n_{s-1})}\,,\qquad \delta e_{p;\,n_1\cdots
n_{s-2}n_{s-1}}\,\eta^{n_{s-1}p}=\partial_m\,\xi^m{}_{n_1\cdots
n_{s-2}}\,\,.
\end{equation}
The comparison of eq. (\ref{ephigauge}) with (\ref{gauge1}) and
(\ref{gauge3}) suggests that the fields $\Phi$ and $D$ of the
triplet are just the symmetric components of the higher--spin
vielbein
\begin{equation}\label{ephiD}
\Phi_{n_1\cdots n_{s}}= s\,e_{(n_s;\,n_1\cdots n_{s-1})}\qquad
D_{n_1\cdots n_{s-2}}=e_{p;\,n_1\cdots
n_{s-2}n_{s-1}}\,\eta^{n_{s-1}p}\,.
\end{equation}
It remains only to identify the field $C$. To this end let us have a
look at the zero torsion condition (\ref{ofc11}). In (\ref{ofc11})
we first symmetrize the index $n$ with $n_1,\ldots,n_{s-1}$ and then
take the trace of $n$ with $m$. In view of eqs. (\ref{ts}) and
(\ref{trnn}) we thus get
\begin{equation}\label{o+e}
(s-1)\,\o_{m;n_1 \cdots
n_{s-1},}{}^{m}=\partial_m\Phi^m{}_{n_1\cdots
n_{s-1}}-(s-1)\,\partial_{(n_{s-1}}\,D_{n_1\cdots
n_{s-2})}-\partial^me_{m;n_1 \cdots n_{s-1}}\,,
\end{equation}
where $\Phi$ and $D$ are defined in (\ref{ephiD}). Comparing
(\ref{o+e}) with (\ref{triplet1}) we see that the triplet field $C$
is actually composed of the trace of the higher--spin connection and
the divergence of the higher--spin vielbein
\begin{equation}\label{C}
C_{n_1 \cdots n_{s-1}}=(s-1)\,\o_{m;n_1 \cdots
n_{s-1},}{}^{m}+\partial^me_{m;n_1 \cdots n_{s-1}}\,.
\end{equation}
We have thus identified the fields of the higher--spin triplet as
components of the higher--spin vielbein and connection of the
frame--like formulation with the relaxed trace constraints.

The comment on the lowest spin fields (\emph{i.e.} the scalar and
the vector) is now in order. As we have already mentioned, these
fields are not contained in the frame--like action (\ref{fract}). In
the case of the even integer spin $s=2k$ the complete trace
component of the vielbein, which could be the scalar field, is
in fact the pure gauge conformal component of the spin two field in the
system. Indeed, as we explained, upon the partial gauge fixing
of the local transformations (\ref{gs1}) the higher--spin vielbein
splits into two completely symmetric tensors (\ref{ephi}). The
maximal trace in all the indices of the rank $s-2$ tensor in
(\ref{ephi}), which might be an independent scalar field, is
identically zero. Hence the scalar field component of the vielbein
is contained only in its rank--$s$ symmetric part (\emph{i.e.} it is
the full trace of $e_{(n_s;\,n_1\cdots n_{s-1})}$) and can be gauged
away by a corresponding residual local transformation
(\ref{ephigauge}). This is similar to the case of gravity where the
trace of the vielbein (or the metric) is a pure gauge scalar
component.

In the case of the odd integer spin $s=2k+1$, as we have explained
earlier, the spin 1 part of the vielbein does not enter the action
(\ref{fract}).
%

As a result, the zero torsion condition (\ref{ofc11}) and its
consequence (\ref{o+e}), which defines the field $C$ (\ref{C}), are
applicable only to the components of the vielbein whose complete
trace in the tangent space indices $n_1,\ldots, n_{s-1}$ is zero
(\emph{i.e.} do not contain the spin 1 field).

To include the scalar and the vector field into the above scheme one
should add to the action (\ref{fract}) corresponding kinetic terms.
As we have mentioned, a systematic way to do this is to use the
zero-forms in the so-called twisted adjoint representation of the
higher--spin algebra. (see e.g. \cite{SHV} for the spin zero case).

To conclude this section we show that the zero torsion condition
(\ref{ofc11}) and the dynamical field equations (\ref{dem}) indeed
imply the equations of motion (\ref{triplet1})--(\ref{triplet3}) of
the triplet higher--spin fields defined by eqs.
(\ref{ephiD})--(\ref{C}). To perform this consistency check, the
following relations between the derivatives of connection components
and the triplet fields (\ref{ephiD}) and (\ref{C})) are useful
\begin{equation}\label{rel1}
2\partial_l\,\omega_{m;n_1\cdots
n_{s-2}}{}^{[l,m]}=-{1\over{s-1}}\,\left(\Box\, D_{n_1\cdots
n_{s-2}}-\partial^m\,C_{mn_1\cdots n_{s-2}}\right)\,,
\end{equation}
\begin{equation}\label{rel2}
\partial_m\,\omega_{(n_1;n_2\cdots
n_{s}),}{}^{m}-\partial_{(n_1}\,\omega^{m;}{}_{n_2\cdots
n_{s}),}{}_{m}={1\over{s(s-1)}}(\Box\Phi_{n_1\cdots
n_{s}}-s\partial_{(n_1}\,C_{n_2\cdots n_s)})\,,
\end{equation}
\begin{eqnarray}\label{rel3}
\partial_l\,\omega_{(n_1;n_2\cdots
n_{s-2})m}{}^{m,l}-\partial_{(n_1}\,\omega^{m;}{}_{n_2\cdots
n_{s-2})l}{}^{l,}{}_{m}&=&{{1}\over{(s-1)(s-2)}}\,\eta^{n_{s-1}n_s}(\Box
\Phi_{n_1\cdots n_{s}}-s\partial_{(n_1}\,C_{n_2\cdots
n_{s})})\nonumber\\
&&\hspace{-50pt} -{{2}\over{(s-1)(s-2)}}\,(\Box D_{n_1\cdots
n_{s-2}}-\partial^{n_{s-1}}\,C_{n_1\cdots n_{s-1}})\,.
\end{eqnarray}
Note that the right hand sides of (\ref{rel1})--(\ref{rel3}) are
proportional to the left hand sides of the triplet field equations
(\ref{triplet2}) and (\ref{triplet3}).

Let us now take the trace of eq. (\ref{dem}) multiplying it with
$\delta_m^b$

\begin{eqnarray}\label{trdem}
&\hspace{-120pt}\delta_m^b\left(\delta^m_{(b}\partial^c\omega^{d;}{}_{n_1\cdots
n_{s-2})[c,d]}+\partial_d\,\omega_{(b;n_1\cdots
n_{s-2})}{}^{[m,d]}+\partial_{(b}\,\omega^{d;}{}_{n_1\cdots
n_{s-2})[d,}{}^{m]}\right)\nonumber\\
\\
&={{2(D+s-4)}\over{s-1}}\,\partial_c\omega_{d;n_1\cdots
n_{s-2}}{}^{[c,d]}+{{s-2}\over{s-1}}\,(\partial_d\,\omega_{(n_1;n_2\cdots
n_{s-2})c}{}^{c,d}-\partial_{(n_1}\,\omega^{d;}{}_{n_2\cdots
n_{s-2})c}{}^{c,}{}_{d})=0\,.\nonumber
\end{eqnarray}
In view of (\ref{rel1})--(\ref{rel3}), eq. (\ref{trdem}) takes the
form
\begin{equation}\label{trdem1}
{(D+s-2)}\,\left(\Box\, D_{n_1\cdots
n_{s-2}}-\partial^m\,C_{mn_1\cdots
n_{s-2}}\right)=\eta^{n_{s-1}n_s}\left(\Box
\Phi_{n_1\cdots n_{s}}-s\partial_{(n_1}\,C_{n_2\cdots
n_{s})}\right)\,.
\end{equation}

Now let us in (\ref{dem}) symmetrize the index $m$ with the indices
$b,n_1,\cdots, n_{s-2}$. The result is
\begin{equation}\label{trdem2}
\,2\eta_{(mb}\,\partial^c\omega^{d;}{}_{n_1\cdots
n_{s-2})[c,d]}+{s\over{s-1}}\,\left(\partial_d\,\omega_{(b;n_1\cdots
n_{s-2}m),}{}^d-\partial_{(b}\,\omega^{d;}{}_{n_1\cdots
n_{s-2}m),}{}_d\right)=0\,.
\end{equation}
To arrive at eq. (\ref{trdem2}) we used the relation
\begin{equation}\label{rel4}
(s-1)\,\omega_{(b;n_1\cdots
n_{s-2}}{}^{d,}{}_{m)}=-\omega_{(b;n_1\cdots n_{s-2}m),}{}^{d}\,
\end{equation}
which is a consequence of the symmetry property (\ref{ts}) of the
connection. By virtue of the relations (\ref{rel1}) and
(\ref{rel2}), eq. (\ref{trdem2}) takes the form
\begin{equation}\label{demsim}
\eta_{(n_1n_2}\,\left(\Box\, D_{n_3\cdots
n_{s})}-\partial^m\,C_{n_3\cdots
n_{s})m}\right)={1\over{(s-1)}}(\Box\Phi_{n_1\cdots
n_{s}}-s\partial_{(n_1}\,C_{n_2\cdots n_s)})\,.
\end{equation}
Comparing eqs. (\ref{trdem1}) and (\ref{demsim}) we conclude that
their left and right hand sides must vanish separately thus
producing the triplet field equations (\ref{triplet2}) and
(\ref{triplet3}).

To recapitulate,  we have shown that, up to a subtlety regarding the
spin--0 and spin--1 field, the higher--spin system described by the
frame--like action (\ref{fract}) for the unconstrained vielbein and
the connection subject to the relaxed trace constraint (\ref{trnn})
is equivalent to the higher--spin triplet. The triplet fields $\Phi$,
$C$ and $D$ have been thus endowed with a geometrical meaning to be
certain components of the higher--spin vielbein and connection. By
singling out these components in the action (\ref{fract}) and
partially solving the zero--torsion condition (\ref{ofc11}) one
should be able to reduce action (\ref{fract}) to the triplet actions
of \cite{Sagnotti:2003qa}.

We shall now extend the results of this section to the AdS background.

\section{Frame--like action for bosonic higher--spin fields in AdS$_D$} \label{flmAdS}
\setcounter{equation}0

The AdS space is described by the vielbein $e^a=dx^m\,e^a_m$ and the
connection $\omega^{ab}=dx^m\,\omega^{ab}_m$ \textbf{which that}
satisfy the following torsion and constant curvature conditions
\begin{equation}\label{adst}
T^a:=de^a+ {\o^{ a}}_{b}\, e^b:=\nabla\,e^a=0\,,
\end{equation}
\be\label{adsR}
R^{ab}(\omega):=d\omega^{ab}+\omega^{a}{}_c\,\omega^{cb}=-\Lambda\,
e^{a} e^{b}\,, 
\ee
where $\nabla=d+\omega$ is the $O(1,D-1)$ covariant differential and
$\Lambda$ is the negative (`cosmological') constant determining the
AdS curvature. The indices from the beginning of the Latin alphabet
now denote the tangent space indices rotated  by the local
$O(1,D-1)$ Lorentz transformations. The indices $m,n,\ldots$ from
the middle of the alphabet denote curved world indices.

The frame--like action for a system of higher--spin fields which
generalizes to AdS the flat space action (\ref{fract}) has the
following form
\begin{eqnarray}\label{AdSactiongf}
& S= \int_{AdS}\,e^{a_1} \ldots e^{{a_{D-3}}}\,\varepsilon_{a_1
\ldots a_{D-3}cdf}\,\left[(\nabla\,e^{b_1
\ldots b_{s-2} c} -{{s-1}\over 2}\, e_k \,\o^{b_1 \ldots b_{s-2} c,\, k})\,
\o_{b_1 \ldots b_{s-2}}{}^{d,\, f}\right.\,\nonumber\\
&\left.\right.\\
&\left. +\Lambda\,{{s(D+s-4)}\over{2(s-1)(D-2)}}\,e^{cb_1\cdots
b_{s-2}}\,e^{d}{}_{b_1\cdots b_{s-2}}\,e^{f}-
\Lambda\,{{(s-2)(s-3})\over{2(D-2)(s-1)}}\,e^{cb_1\cdots
b_{s-4}j}{}_j\,e^{d}{}_{b_1\cdots
b_{s-4}i}{}^i\,e^{f}\right]\,.\nonumber
\end{eqnarray}

 We observe that, apart from covariantized derivatives,
  the action (\ref{AdSactiongf}) differs from
the flat space action (\ref{fract}) by the last two mass--like terms
proportional to the AdS curvature. Note that the last term in
(\ref{fract}) contains the trace of the higher--spin vielbein. As is
well known, in AdS space such terms are required to keep a number of
physical states of the higher--spin field equal to that of the
massless field, \emph{i.e.}
 to preserve gauge symmetries. Hence,
the coefficients in front of these terms are fixed by the
requirement of the invariance of this action under the gauge
transformations of the higher--spin vielbein and connection whose
form we shall discuss in the next two subsections.

Here we only note that, as in the flat case, the higher--spin
vielbein is unconstrained (modulo the subtlety that it does not
contain the spin--1 field, as was discussed in detail in Section 2),
while the variation of the action (\ref{AdSactiongf}) with respect
to the higher--spin connection produces the zero torsion condition
\begin{equation}\label{ofc11ads}
T^{a_1 \cdots a_{s-1}}=0\qquad \Longleftrightarrow
\qquad (s-1)\o_{[n;}{}^{a_1
\cdots
a_{s-1},b}\,e_{m]b}=\nabla_{[m}\,e_{n];}{}^{a_1\cdots a_{s-1}}\,,
\end{equation}
provided that the higher--spin connection obeys the relaxed traceless
condition
\begin{equation}\label{trads}
\eta_{a_1b}\,\omega^{a_1a_2\cdots a_{s-1},b}=0\,.
\end{equation}
The dynamical field equation of the higher--spin vielbein in AdS
gets modified by the contribution of the terms proportional to the
AdS curvature $\Lambda$ and acquires the form
\begin{eqnarray}\label{demads}
&\left(\nabla_n \o_{r;}{}_{(a_1 \cdots a_{s-2}}{}^{c,\,
d}-\Lambda\,{{s(D+s-4)}\over{(s-1)(D-2)}}\,e^{d}_n\,e_{r;(a_1\cdots
a_{s-2}}{}^{c}\right.\hspace{200pt}\\
&
\hspace{150pt}
\left.+\Lambda\,{{(s-2)(s-3})\over{2(D-2)(s-1)}}\,e^{d}_n\,e_{r;(a_3\cdots
a_{s-2}}{}^{c}\,\eta_{a_1a_2}\right)\,e_{b)}^{[m}e_{c}^{n}e_{d}^{r]}=0\,.\nonumber
\end{eqnarray}

\subsection{Fronsdal case}\label{adsbf}
The frame-like action for irreducible massless fields in $AdS_D$
originally proposed in \cite{LV} was, in fact, the first action for
symmetric massless fields in  $D>4$.\footnote{The metric-like
formulation of Fronsdal  was originally proposed in
\cite{fronsdal78,fronsdal78a} for the case of $D=4$. It turns out
that the coefficients in front of different terms of the action are
independent of $D$ in Minkowski space but those of mass-like terms
are  $D$-dependent in $AdS_D$.}
 The action constructed in
\cite{LV} is manifestly gauge invariant due to the use of higher
connections called extra fields, which however do not contribute to
the free field equations. A version of this approach which is
manifestly $o(2,D-1)$ (rather than $o(1,D-1)$) invariant was later
proposed in
\cite{5d}. In Section \ref{ADSC} we shall demonstrate how this approach works
in the case of the relaxed higher--spin multiplets being the main
subject of this paper. In this section we shall use an alternative
approach in which the gauge symmetry is not manifest but the
analysis is simpler since the set of gauge fields is free from the
extra fields.

Imposing the conventional traceless conditions on the higher--spin
vielbein and connection (as above we distinguish between the
traceless and traceful quantities by putting tildes on the former)
\begin{equation}\label{traceless}
\eta_{a_1a_2}\,\tilde e^{a_1a_2\cdots
a_{s-1}}=0,\quad \eta_{a_1a_2}\,\tilde \omega^{a_1a_2\cdots
a_{s-1},b}=0,
\end{equation}
the action (\ref{AdSactiongf}) reduces to
\begin{eqnarray}\label{AdSactiongf1}
& S= \int_{AdS}\,e^{a_1} \ldots e^{{a_{D-3}}}\,\varepsilon_{a_1
\ldots a_{D-3}cdf}\,\left[(\nabla\,\tilde e^{b_1
\ldots b_{s-2} c} -{{s-1}\over 2}\, e_k \,\tilde\o^{b_1 \ldots b_{s-2} c,\, k})\,
\tilde\o_{b_1 \ldots b_{s-2}}{}^{d,\, f}\right.\,\nonumber\\
&\left.\right.\\
&\left.\hspace{200pt}
+\Lambda\,{{s(D+s-4)}\over{2(s-1)(D-2)}}\,\tilde e^{cb_1\cdots
b_{s-2}}\,\tilde e^{d}{}_{b_1\cdots
b_{s-2}}\,e^{f}\right]\,.\nonumber
\end{eqnarray}
It is invariant under the following gauge transformations of the
higher--spin vielbein and connection
\begin{equation}\label{gsads11}
\delta \,\tilde e^{a_1 \ldots a_{s-1}}=\nabla\tilde\xi^{a_1 \ldots a_{s-1}}-(s-1)\,e^c\,\tilde\xi^{a_1 \ldots
a_{s-1},b}\,\eta_{bc}\,,
\end{equation}
\begin{eqnarray}\label{gsads12}
&\hspace{-20pt}\delta\, \tilde\omega^{a_1 \ldots a_{s-1},b}=
\nabla\,\tilde\xi^{a_1
\ldots a_{s-1},b}-(s-2)\,e^c\,\tilde\xi^{a_1
\ldots a_{s-1},bd}\,\,\eta_{cd}-\Lambda\,(e^b\,\tilde\xi^{a_1 \ldots a_{s-1}}-e^{(a_1}\,\tilde\xi^{a_2
\ldots a_{s-1})b})\,,
\end{eqnarray}
where the parameters $\tilde\xi^{a_1 \ldots a_{s-1}}$ and $\tilde
\xi^{a_1
\ldots a_{s-1},b}$ are traceless and the parameter $\tilde\xi^{a_1
\ldots a_{s-1},bd}$ satisfies the following trace conditions \footnote{The transformations
(\ref{gsads12}) and the trace conditions (\ref{trelate}) can be
obtained from the $O(2,D-1)$--covariant expressions
(\ref{gt})--(\ref{Vcontruction2}) of Section
\ref{ADSC} in the standard gauge (\ref{vgauge}).}
\begin{eqnarray}\label{trelate}
\hspace{-30pt}
 \eta_{a_1a_2}\,\tilde\xi^{a_1a_2\cdots
a_{s-1},b_1b_2}=\frac{2\Lambda}{(s-1)(s-2)}\,\tilde\xi^{a_3\cdots
a_{s-1}b_1b_2},\nonumber\\
\\
\quad
\eta_{a_1b_1}\,\tilde\xi^{a_1a_2\cdots
a_{s-1},b_1b_2}=-\frac{\Lambda}{s-1}\,\tilde\xi^{a_2\cdots
a_{s-1}b_2} .\nonumber
\end{eqnarray}
Note that the second relation is a consequence of the first one by
virtue of the Young symmetry properties of $\tilde\xi^{a_1
\ldots a_{s-1},bd}$.

In the flat limit $\Lambda \rightarrow 0$, eqs.
(\ref{gsads11})--(\ref{trelate}) reduce to the corresponding gauge
transformations discussed in Subsection \ref{fronsdal}.

Action (\ref{AdSactiongf1}) describes in AdS space the dynamics of a
single massless field of spin $s$. Upon solving $\tilde
\omega^{a_1\cdots a_{s-1},b}$ in terms of $\tilde e^{a_1\cdots
a_{s-1}}$ and partially fixing local higher--spin symmetry
(\ref{gsads11}), (\ref{gsads12}) the action (\ref{AdSactiongf1})
gives the generalization to any dimension of the $AdS_4$ Fronsdal
action \cite{fronsdalAdS} for the double traceless field
$\phi^{a_1\cdots a_s}:=e^{m(a_1}\,\tilde e_{m;}{}^{a_2\cdots a_s)}$.

\subsection{Triplet case}\label{tripletcase2}
If the full traceless condition is not imposed, the higher--spin
connection only satisfies the relaxed trace condition (\ref{trads})
while the higher--spin vielbein remains unconstrained. Then the
action (\ref{AdSactiongf}) describes in AdS space a system of free
massless fields of descending spins  $s-2, s-4,
\ldots , 3$ or 2 depending on whether $s$ is odd or even. The
analysis and the proof is the same as in the flat case (see
Subsection \ref{tripletcase1}). The only difference is that the
gauge transformations of the higher--spin vielbein and connection
which leave the action (\ref{AdSactiongf}) invariant and which have
the same form as eqs. (\ref{gsads11}) and (\ref{gsads12})  contain
the unconstrained parameter $\xi^{a_1 \ldots a_{s-1}}$, $\xi^{a_1
\ldots a_{s-1},b}$ satisfies the relaxed traceless condition
\begin{equation}\label{rtads1}
\xi^{a_1 \ldots a_{s-1},b}\,\eta_{a_1b}=0\,,
\end{equation}
while the parameter $\xi^{a_1
\ldots a_{s-1},bd}$  is subject to the following
relaxed constraint
\begin{eqnarray}\label{xibd}
(s-1)\, \eta_{bc}\,\xi^{a_1
\ldots a_{s-2}b,cd}=\Lambda \,(\,\eta^{d(a_1}\,
\xi^{a_2\ldots a_{s-2})b}{}_b-\xi^{a_1\ldots a_{s-2}d})
\end{eqnarray}
instead of being traceless as in the flat space case (see eq.
(\ref{trmn})) or `partially' traceless in the Fronsdal AdS case (see
eq. (\ref{trelate})). We shall also use the following consequences
of (\ref{xibd}) and of the Young symmetry  $Y(s-1,2)$ of
$\xi^{a_1\ldots a_{s-1},bc}$
\begin{eqnarray}
 \eta_{bc}\,\xi^{a_1\ldots
a_{s-1},bc}=\Lambda\,(\xi^{a_1\ldots a_{s-1}}-\eta^{(a_1a_2}\,
\xi^{a_3\ldots a_{s-1})b}{}_b)\,,\hspace{90pt}\label{xibd1}\\
\eta_{bc}\,\xi^{bc(a_1\ldots
a_{s-3},a_{s-2})d}={{2\Lambda}\over{(s-1)(s-2)}}\,(\xi^{a_1\ldots
a_{s-2}d}-\eta^{d(a_1}\,
\xi^{a_2\ldots a_{s-2})b}{}_b)\label{xibd2}\,.
\end{eqnarray}

Let us now identify the AdS higher--spin triplet in terms of
components of the higher--spin vielbein and connection. In the AdS
space the bosonic higher--spin triplet is defined (in our notation
and convention) by the following equations
\cite{Francia:2002pt,Sagnotti:2003qa}
\begin{equation}\label{triplet1ads}
C_{n_1\cdots n_{s-1}}=\nabla_m\Phi^m{}_{n_1\cdots
n_{s-1}}-(s-1)\,\nabla_{(n_{s-1}}\,D_{n_1\cdots n_{s-2})}\,,
\end{equation}
\begin{eqnarray}\label{triplet2ads}
\Box\,\Phi_{n_1\cdots n_s}=s\,\nabla_{(n_{s}}\,C_{n_1\cdots
n_{s-1})}\,+\Lambda\,\left[(s-(s-2)(D+s-3))\,\Phi_{n_1\cdots
n_s}\right.\\
&
\left. {\hspace{-120pt}}+2s(s-1)\,g_{(n_1n_2}(\Phi_{n_3\cdots n_s)ml}\,g^{ml}
-4D_{n_3\cdots n_s)}) \right],\nonumber
\end{eqnarray}
\begin{eqnarray}\label{triplet3ads}
\Box\,D_{n_1\cdots n_{s-2}}=\nabla_m\,C^m{}_{n_1\cdots
n_{s-2}}-\Lambda\left[(s(D+s-2)+6)\,D_{n_1\cdots n_{s-2}}\right.\\
&{\hspace{-170pt}}
\left.-4\,\Phi_{n_1\cdots
n_{s-2}ml}\,g^{ml}-(s-2)(s-3)\, g_{(n_1n_2}\,D_{n_3\cdots
n_{s-2})ml}\,g^{ml}\right]\nonumber
\,,
\end{eqnarray}
where $\Box:=\nabla_m\,\nabla^m$ and
$g_{mn}=e^a_m\,e^b_{n}\,\eta_{ab}$ is the AdS metric.

 The equations (\ref{triplet1})--(\ref{triplet3}) are invariant
under the following gauge transformations
\begin{equation}\label{gauge1ads}
\delta \Phi_{n_1\cdots n_s}=s\,\nabla_{(n_{s}}\,\xi_{n_1\cdots
n_{s-1})}
\end{equation}
\begin{equation}\label{gauge3ads}
\delta D_{n_1\cdots
n_{s-2}}=\nabla_m\,\xi^m{}_{n_1\cdots n_{s-2}}\,,
\end{equation}
\begin{eqnarray}\label{gauge2ads}
\delta C_{n_1\cdots n_{s-1}}&=&\Box\,\xi_{n_1\cdots n_{s-1}}-\Lambda (D+s-3)\,(s-1)\,\xi_{n_1\cdots
n_{s-1}}\\
&&\hspace{170pt}+(s-1)(s-2)\Lambda\,g_{(n_1n_2}\,\xi_{n_3\cdots
n_{s-1})lm}\,g^{lm}\,,\nonumber
\end{eqnarray}
where the parameter $\xi_{n_1\cdots n_{s-1}}$ is completely
symmetric and traceful.

As in the flat case, we can identify the fields $\Phi_{n_1\cdots
n_s}$ and $D_{n_1\cdots n_{s}}$ with the completely symmetric part
and a trace part of the higher--spin vielbein $e_{m;}{}^{a_1\cdots
a_{s-1}}$, respectively,
\begin{equation}\label{ephiDads}
\Phi_{n_1\cdots n_{s}}= s\,e_{(n_s;\,n_1\cdots n_{s-1})}\qquad
D_{n_1\cdots n_{s-2}}=e_{p;\,n_1\cdots
n_{s-2}n_{s-1}}\,g^{pn_{s-1}}\,,
\end{equation}
where the tangent space indices of the higher--spin vielbein have
been converted into the `curved' world indices with the use of the
AdS vielbein $e_{m}^a$. The field $C_{n_1\cdots n_{s-1}}$ is then
identified by analyzing the zero torsion condition (\ref{ofc11ads})
and has the form similar to (\ref{C}), namely
\begin{equation}\label{Cads}
C_{n_1 \cdots n_{s-1}}=(s-1)\,\o_{m;n_1 \cdots
n_{s-1},}{}^{m}+\nabla^m\,e_{m;n_1 \cdots n_{s-1}}\,.
\end{equation}
As in the flat space case, one can show that the gauge
transformations (\ref{gauge1ads})--(\ref{gauge2ads}) and the
equations of motion (\ref{triplet1ads})--(\ref{triplet3ads}) of the
triplet fields follow, respectively, from the transformations
(\ref{gsads11})--(\ref{gsads12}) and the equations of motion
(\ref{ofc11ads})--(\ref{demads}). For instance, the last two terms
in the variation of the field $C$ (\ref{gauge2ads}) come from the
terms of the gauge variation of the higher--spin connection which are
proportional to $\Lambda$  (see eqs. (\ref{gsads12}) and
(\ref{xibd1})).

By singling out the fields (\ref{ephiDads}) and (\ref{Cads})
 in the action (\ref{AdSactiongf}) and partially solving the
zero--torsion condition (\ref{ofc11ads}) one should be able to
reduce the action (\ref{AdSactiongf}) to the AdS triplet actions of
\cite{Sagnotti:2003qa} for $s\geq 2$. As has been already explained
in the case of flat
space--time, the scalar and the vector fields are not part of the
triplet spectrum in our formulation, but they can be included into
the model by adding corresponding terms to the AdS action
(\ref{AdSactiongf}).

\section{Frame--like action for fermionic higher--spin fields in flat space--time}
\label{ferm}
\setcounter{equation}0

The frame-like formulation of irreducible higher--spin fermions was
originally proposed  in $D=4$ Minkowski space  in \cite{V80,DA}
solely in terms of the frame-like fields, then extended to $AdS_4$
using the formalism of two-component spinors in \cite{V87}, where
also the extra field connections were introduced, and then to
$AdS_D$ with any $D$ in \cite{Vasiliev:1987tk}. The difference
compared to the bosonic case is that the fermionic field equations
are of the first order and hence the free action does not contain
auxiliary fields.

The flat space higher--spin field strengths (curvatures) are of the
same form as in the bosonic case \cite{Vasiliev:1987tk}
\be
\label{fs}
R_{a_1\ldots a_{s-{3\over 2}}\,,b_1\ldots b_{t}}= d\psi_{a_1\ldots
a_{s-{3\over 2}}\,,b_1\ldots b_{t}}-
(s-t-\frac{3}{2})\,e^c\psi_{a_1\ldots a_{s-{3\over 2}}\,,b_1\ldots
b_{t} c}
\ee
where $\psi^\alpha_{a_1\ldots a_{s-{3\over 2}}\,,b_1\ldots b_{t}}=
dx^n\psi^\alpha_{n\,; a_1\ldots a_{s-{3\over 2}}\,,b_1\ldots b_{t}}$
is a one-form connection (with respect to the index $n$) and a rank
$s-\frac{3}{2} +t$ tensor-spinor ( $0\leq t\leq s-\frac{3}{2}$ and
$\alpha$ being a (usually implicit) index associated with a spinor
representation of $Spin(1,D-1)$\footnote{In a generic D--dimensional
space--time the spinors are of the Dirac type. In even dimensions
one can restrict spinors to be Weyl and in certain dimensions,
\emph{e.g.} $D=3,4,6,10$ and $11$, one can consider Majorana or
symplectic Majorana tensor--spinors. Note, however, that in the
even--dimensional AdS spaces the Weyl condition cannot be imposed
due to the presence of mass--like terms (see Section
\ref{AdSftriplets}). In the case of the Dirac and Weyl spinors the
actions which we consider below implicitly contain the hermitian
conjugate part, which we shall skip for brevity. The addition of the
hermitian conjugate part makes the first--order Lagrangian for Dirac
fermions real exactly (\emph{i.e.} not only up to a total
derivative).\label{majorana}}). The field strengths (\ref{fs}) are
manifestly invariant under the gauge transformations
\be
\label{fgtr}
\delta\psi_{a_1\ldots a_{s-{3\over 2}}\,,b_1\ldots b_{t}}=
d\xi_{a_1\ldots a_{s-{3\over 2}}\,,b_1\ldots b_{t}}-
(s-t-\frac{3}{2})\,e^c\xi_{a_1\ldots a_{s-{3\over 2}}\,,b_1\ldots
b_{t} c}\,.
\ee

As in the bosonic case, the symmetry properties of the fermionic
higher--spin fields and of the gauge parameters are governed by the
Young tableaux. The tensor--spinor fields with $t\geq 1$ are of
extra type and will not participate in the description of the free
higher-spin fermionic system. They play an important role, though,
in the construction of the consistent interacting higher-spin theory
\cite{fvnp,Alkalaev:2002rq,bciv}.

Let us consider the following first order action for the fermionic
higher--spin field
\be \label{feract} S= i\int_{M^D}\,e^{a_1} \ldots
e^{{a_{D-3}}}\,\varepsilon_{a_1 \ldots a_{D-3}pqr} (
\bar{\psi}_{d_1\ldots d_{s-{3\over 2}}} \gamma^{pqr} d\psi^{d_1\ldots
d_{s-{3\over 2}}} +\alpha\, \bar{\psi}_{d_1\ldots d_{s-{5\over
2}}}{}^p\gamma^{q}
\,d\,\psi^{d_1
\ldots d_{s-{5\over 2}} }{}^r )\,,
 \ee
where $\alpha$ is a constant parameter and
\be
\gamma^{pqr} = \frac{1}{6}
\left ((\gga^p \gga^q \gga^r - \gga^q \gga^p \gga^r)
+ \mbox{two cyclic permutations of\quad} p,q,r \right ).
\ee
The value of the parameter $\alpha = -6 (s-\frac{3}{2})$ is fixed by
requiring the invariance of the action (\ref{feract}) under the
gauge transformations (\ref{fgtr}). Indeed, the action is manifestly
gauge invariant provided that it can be reformulated in the
following form
\be \label{sr} S= i\int_{M^D}\,e^{a_1} \ldots
e^{{a_{D-3}}}\,\varepsilon_{a_1 \ldots a_{D-3}pqr} (
\bar{\psi}_{d_1\ldots d_{s-{3\over 2}}} \gamma^{pqr} R^{d_1\ldots d_{s-{3\over 2}}}
+\alpha \,\bar{\psi}_{d_1\ldots d_{s-{5\over 2}}}{}^p\gamma^{q}
\,R^{d_1
\ldots d_{s-{5\over 2}} }{}^r )
 \, \ee
 and (up to a total derivative) as
\be \label{sbr} S= i\int_{M^D}\,e^{a_1} \ldots
e^{{a_{D-3}}}\,\varepsilon_{a_1 \ldots a_{D-3}pqr} (
\bar{R}_{d_1\ldots d_{s-{3\over 2}}} \gamma^{pqr}
\psi^{d_1\ldots d_{s-{3\over 2}}}
+\alpha\, \bar{R}_{d_1\ldots d_{s-{5\over 2}}}{}^p\gamma^{q}
\,\psi^{d_1
\ldots d_{s-{5\over 2}} }{}^r )
 \,. \ee
In view of the definition of the curvature (\ref{fs}) with $t=0$,
the equivalence of (\ref{feract}) and (\ref{sr}),  (\ref{sbr})
requires that the connections $\psi_{a_1\ldots a_{s-{3\over
2}}\,,b}$ and $\bar{\psi}_{a_1\ldots a_{s-{3\over 2}}\,,b}$ do not
contribute, respectively, to the action (\ref{sr}) and (\ref{sbr}).
Note that in the form (\ref{sr}) the action is manifestly invariant
under the gauge variations of $\psi$ (\ref{fgtr}) (with $t=0,1$),
while in the form (\ref{sbr}) it is manifestly gauge invariant under
the gauge variation of the Dirac conjugate connections $\bar\psi$.

Thus, the possibility of rewriting  the same action (\ref{feract})
both in the form (\ref{sr}) and (\ref{sbr}) imposes constraints on
the connection $\psi^{a_1\ldots a_{s-\frac{3}{2}}\,,b}$ and the
corresponding gauge parameter $\xi^{a_1\ldots
a_{s-\frac{3}{2}}\,,b}$. To find these constraints we use the
identity (\ref{idep}) to present the $\psi^{a_1\ldots
a_{s-\frac{3}{2}}\,,b}$--dependent part of the action (\ref{sr}) in
the form
\bee
&{}& X= -\frac{s-\frac{3}{2}}{D-2}\,i\,\int_{M^D}\,e^{a_1} \ldots
e^{{a_{D-2}}}\,\varepsilon_{a_1 \ldots a_{D-3}pq}
\Big (3\bar{\psi}_{a_1\ldots a_{s-\frac{3}{2}}}( \gamma^{p}
\gamma^{q} \gamma^{r}
-2 \eta^{qr} \gamma^p )\psi^{a_1\ldots a_{s-\frac{3}{2}}}{}_{,\,r}\nn\\
&\\
\ls&+&\alpha\,\Big ( \bar{\psi}_{a_1\ldots a_{s-\frac{5}{2}}}{}^p\gamma^{q}
\,\psi^{a_1 \ldots a_{s-\frac{5}{2}} }{}^r{}_{,r}
+ \bar{\psi}_{a_1\ldots a_{s-\frac{5}{2}}}{}^q\gamma^{r}
\,\psi^{a_1 \ldots a_{s-\frac{5}{2}} }{}^p{}_{,r}
- \f{1}{s-\frac{3}{2}} \bar{\psi}_{a_1\ldots
a_{s-\frac{3}{2}}}{}\gamma^{p}
\,\psi^{a_1 \ldots a_{s-\frac{3}{2}}\,,q}\Big )\Big )
 \,\nonumber
\eee

Setting
\be
\label{ab}
\alpha = -6 (s-\frac{3}{2})\,,
\ee
we find that $X$ vanishes provided that
\be
\label{fredc}
\gga_b \,\psi^{a_1 \ldots a_{s-\frac{3}{2}}\,,b}=0\q
\psi^{a_1 \ldots a_{s-\frac{5}{2}}b,}{}_b=0\,.
\ee
These conditions provide a fermionic analog of the relaxed traceless
condition (\ref{trnn}). Note that the fermionic higher--spin
vielbein $\psi_{n_1 \ldots n_{s-\frac{3}{2}}}$ remains
unconstrained.

The action (\ref{feract}) or (\ref{sr}) and (\ref{sbr}) is invariant
under the gauge transformations (\ref{fgtr}) (with $t=0,1$) provided
that the gauge parameters $\xi^{a_1 \ldots a_{s-\frac{3}{2}}\,,b}$
satisfy the constraints analogous to (\ref{fredc})
\be
\label{fredc1}
\gga_b\, \xi^{a_1 \ldots a_{s-\frac{3}{2}}\,,b}=0\q
\xi^{a_1 \ldots a_{s-\frac{5}{2}}b,}{}_b=0\,.
\ee
The variation of the action (\ref{feract}) with respect to $\psi$
produces the following equations of motion
\begin{eqnarray}\label{ffe}
\frac{1}{s-\frac{3}{2}}\,\gamma^{mqr}\partial_r\,\psi_{q;a_1\cdots
a_{s-\frac{3}{2}}}= \gamma^m\,\partial^r\,
\psi_{(a_1;a_2\cdots a_{s-\frac{3}{2}})r}-\gamma^q\,\partial^r\,\psi_{q;r(a_2\cdots
a_{s-\frac{3}{2}}}\,\delta^m_{a_1)}\nonumber\\
-\gamma^r\,\partial_r\,\psi_{(a_1;a_2\cdots
a_{s-\frac{3}{2}})}{}^{m}
+\gamma^r\,\partial_r\,\psi^{p;}{}_{p(a_2\cdots
a_{s-\frac{3}{2}}}\delta^{m}_{a_1)}\\
-\gamma^m\,\partial_{(a_1}\,\psi^{q;}{}_{a_2\cdots
a_{s-\frac{3}{2}})q}+\gamma^q\,\partial_{(a_1}\,\psi^{m;}{}_{a_2\cdots
a_{s-\frac{3}{2}})q}\,.\nonumber
\end{eqnarray}

\subsection{Fang--Fronsdal case}
Let us now consider the case in which the fermionic higher--spin
vielbein, the connections  and the gauge parameters are required to
be gamma--transversal (or gamma--traceless) and hence traceless
\emph{in all} tangent space indices
\bee\label{ffrcon}
\gga^c \tilde\psi_{a_1\ldots a_{s-\frac{3}{2}}\,,b_1\ldots b_{t} c}=0\q
\gga^c \tilde\psi_{a_1\ldots a_{s-\frac{5}{2}} c\,,b_1\ldots
b_{t}}=0,\\
\gga^c \tilde\xi_{a_1\ldots a_{s-\frac{3}{2}}\,,b_1\ldots b_{t} c}=0\q
\gga^c \tilde\xi_{a_1\ldots a_{s-\frac{5}{2}} c\,,b_1\ldots b_{t}}=0.
\eee
As we shall now demonstrate, in this case the action (\ref{feract})
is the frame--like counterpart \cite{V80,DA} of the Fang--Fronsdal action
\cite{fronsdal78a} for a single fermionic field of a half--integer spin $s$ in flat
space.

The gamma--tansversal higher--spin vielbein $\tilde\psi^\alpha_{m;\,
n_1
\cdots n_{s-\frac{3}{2}}}$ contains the
irreducible Lorentz  tensor-spinors \footnote{As in the bosonic
case, in flat space--time we do not distinguish between symmetric
tangent space indices $a,b,\ldots$ and world indices $m,n, \ldots$.
The latter are separated from the former by `;'.} described by the
following gamma-transversal Young tableaux
\be
\begin{picture}(6,7)(0,0)
\multiframe(0,0)(7.5,0){1}(7,7){}
\end{picture}\
\otimes
\begin{picture}(50,12)(0,0)
\multiframe(0,0)(13.5,0){1}(30,7){}\put(33,1){\tiny{$ s-\frac{3}{2}$}}
\end{picture}\ =
\begin{picture}(50,12)(0,0)
\multiframe(0,0)(13.5,0){1}(30,7){}\put(33,1){\tiny{$ s-\frac{1}{2}$}}
\end{picture}\oplus
\begin{picture}(50,12)(0,0)
\multiframe(0,0)(13.5,0){1}(30,7){}\put(33,1){\tiny{$ s-\frac{3}{2}$}}
\end{picture}\oplus
\begin{picture}(55,12)(0,0)
\multiframe(0,0)(13.5,0){1}(30,7){}\put(33,1){\tiny{$ s-\frac{5}{2}$}}
\end{picture}\oplus
\begin{picture}(60,15)(-5,5)
\multiframe(0,0)(13.5,0){1}(7,7){}\put(9,1){{\tiny $ 1$}}
\multiframe(0,7.5)(13.5,0){1}(35,7){}\put(38,9){\tiny{$ s-\frac{3}{2}$}}
\end{picture}\,\,.
\label{ef}\ee
The first tableau of length $s$ on the right hand side of (\ref{ef})
describes the totally symmetric and gamma--transversal part $
\tilde \psi_{(m;\, n_1
\cdots n_{s-\frac{3}{2}}) }$, the
second and third tableaux of the length $s-\frac{3}{2}$ and
$s-\frac{5}{2}$, respectively, corresponds to the contractions
$\gamma^m \tilde \psi_{m;\, n_1
\cdots n_{s-\frac{3}{2}}} $ and $\eta^{mk}\tilde \psi_{m;\, n_1
\cdots n_{s-\frac{5}{2}}k}{}$, respectively.  The hook tableau
corresponds to the irreducible (gamma--transversal) part of $\tilde
\psi$ that satisfies $\tilde \psi_{(m;\, n_1\cdots
n_{s-\frac{3}{2}})}=0$.

In virtue of the gauge transformations of the higher--spin vielbein
(with gamma--traceless parameters)
\be\label{deviel}
\delta\,\tilde\psi^{n_1\cdots n_{s-\frac{3}{2}}}=d\tilde\xi^{n_1\cdots n_{\s-1}}
- (s-\frac{3}{2})\,e^m\,\tilde \xi^{n_1\cdots
n_{s-\frac{3}{2}},b}\eta_{mb}\,,
\ee
the hook part of the higher--spin vielbein field can be gauge fixed
to zero by the appropriate choice of the parameter
$\tilde\xi^{a_1\cdots a_{s-\frac{3}{2}},b}$. As a result, the
remaining part of the vielbein amounts to the combination of three
totally symmetric gamma-transversal tensor--spinors of rank
$s-\frac{1}{2}$, $s-\frac{3}{2}$ and $s-\frac{5}{2}$ which are
equivalent to the Fang--Fronsdal symmetric tensor--spinor field
$\Psi^\alpha_{n_1\cdots n_{s-\frac{1}{2}}}$ that satisfies the
triple gamma--traceless condition
\be
\gamma^l\,\gamma^m\,\gamma^p\,\Psi_{lmp n_1\cdots n_{s-\frac{5}{2}}}
=\eta^{lm} \gga^p \Psi_{lmp n_1\cdots n_{s-\frac{5}{2}}}=0\,.
\ee

The remaining local symmetry is the gauge invariance
 of the Fang--Fronsdal metric--like formulation in flat
space--time with the completely symmetric gamma--transversal
tensor--spinor parameter $\tilde\xi_{m_1
\cdots m_{s-\frac{3}{2}}}$
\be
\gamma^n\, \tilde\xi_{m_1 \cdots m_{s-\frac{5}{2}} n} =0\,.
\ee
Thus, the action (\ref{feract}) or (\ref{sr}) with the fields
restricted by the conditions (\ref{ffrcon}) is equivalent
 \cite{V80,DA,Vasiliev:1987tk} to the Fang-Fronsdal action.

\subsection{Triplet case}
Let us now consider the case in which the fermionic higher--spin
vielbein $\psi^{a_1\ldots a_{s-\frac{3}{2}}}$ and the gauge
parameter $\xi^{a_1\ldots a_{s-\frac{3}{2}}}$ are unconstrained
while the parameter $\xi^{a_1\ldots a_{s-\frac{3}{2}},c}$ of the
gauge transformation (\ref{fgtr}) (for $t=0$) is constrained by the
relaxed conditions
\be
\label{fredcxi}
\gga_b\, \xi^{a_1 \ldots a_{s-\frac{3}{2}}\,,b}=0\q
\xi^{a_1 \ldots a_{s-\frac{5}{2}}b,}{}_b=0 \quad \Longrightarrow \quad
[\gga^c\,,\gga^d] \xi_{a_1\ldots a_{\s-2} c\,, d}=0\,.
\ee

In order to figure out what is  the field content of the model in
this case we observe that, the one-form fermionic field
$\psi_{a_1\ldots a_{s-\frac{3}{2}}}=dx^m\,\psi_{m;a_1\ldots
a_{s-\frac{3}{2}}}$ is composed of the tensor-spinors characterized by the
following \emph{unrestricted} (\emph{i.e.} gamma--traceful) Young
tableaux
\be
\begin{picture}(6,7)(0,0)
\multiframe(0,0)(7.5,0){1}(7,7){}
\end{picture}\
\otimes
\begin{picture}(50,12)(0,0)
\multiframe(0,0)(13.5,0){1}(30,7){}\put(33,1){\tiny{$ {s-\frac{3}{2}}$}}
\end{picture}\ =
\begin{picture}(50,12)(0,0)
\multiframe(0,0)(13.5,0){1}(30,7){}\put(33,1){\tiny{$ {s-\frac{1}{2}}$}}
\end{picture}\oplus
\begin{picture}(60,15)(-5,5)
\multiframe(0,0)(13.5,0){1}(7,7){}\put(9,1){{\tiny $ 1$}}
\multiframe(0,7.5)(13.5,0){1}(35,7){}\put(38,9){\tiny{$ {s-\frac{3}{2}}$}}
\end{picture}\,\,.
\label{ef1}\ee
On the other hand, the parameter $\xi_{\a_1\ldots a_{\s-1}\,,b}$ of
the Stueckelberg gauge symmetry  that satisfies (\ref{fredcxi}) has
the following components
\be
\begin{picture}(60,15)(-5,5)
\multiframe(0,0)(13.5,0){1}(7,7){}\put(9,1){{\tiny $ 1$}}
\multiframe(0,7.5)(13.5,0){1}(35,7){}\put(38,9){\tiny{${s-\frac{3}{2}}$}}
\end{picture}\,\,\Big / (
\begin{picture}(50,12)(0,0)
\multiframe(0,0)(13.5,0){1}(30,7){}\put(33,1){\tiny{${s-\frac{3}{2}}$}}
\end{picture}\oplus
\begin{picture}(55,12)(0,0)
\multiframe(0,0)(13.5,0){1}(30,7){}\put(33,1){\tiny{${s-\frac{5}{2}}$}}
\end{picture})\,\,
\label{ef2}\ee
where the subtracted  (factored out) tensors take into account the
two conditions (\ref{fredcxi}). As a result, we find that, upon
gauge fixing to zero the pure gauge part of $\psi_{a_1\ldots
a_{s-\frac{3}{2}}}$ associated with the Stueckelberg symmetry, the
remaining components of the fermionic field are described by the sum
of the following unrestricted Young tableaux
\be
\begin{picture}(50,12)(0,0)
\multiframe(0,0)(13.5,0){1}(30,7){}\put(33,1){\tiny{$s-\frac{1}{2}$}}
\end{picture}\oplus
\begin{picture}(50,12)(0,0)
\multiframe(0,0)(13.5,0){1}(30,7){}\put(33,1){\tiny{$s-\frac{3}{2}$}}
\end{picture}\oplus
\begin{picture}(55,12)(0,0)
\multiframe(0,0)(13.5,0){1}(30,7){}\put(33,1){\tiny{$s-\frac{5}{2}$}}
\end{picture}\,\,.
\label{phf}\ee
Each term in (\ref{phf}) describes \emph{unconstrained} totally
symmetric spinor--tensors of ranks $s-\frac{1}{2}$, $s-\frac{3}{2}$
and $s-\frac{5}{2}$, respectively. Decomposing this set of fields
into Lorentz irreducible gamma-traceless components, we get the set
of Fang-Fronsdal massless fields of the half--integer spins
descending from $s$ down to $3/2$. Note that, analogously to the
fields of spin one and zero in the bosonic case, the spin--1/2 field
(being a zero--form) is not described by the action (\ref{feract})
and should be treated separately.

Up to this subtlety, the field content of the model under
consideration is the same as that of the fermionic higher--spin
triplets
\cite{Francia:2002pt,Sagnotti:2003qa}. Since both models describe
free fields, there should be a relation between them.

To find this relation let us look at the form of the equations and
gauge transformations which define the triplet of unconstrained
fermionic higher--spin fields $\Psi_{m_1\cdots m_{s-\frac{1}{2}}}$,
$\chi_{m_1\cdots m_{s-\frac{3}{2}}}$ and $\lambda_{m_1\cdots
m_{s-\frac{5}{2}}}$ in flat space--time
\cite{Francia:2002pt,Sagnotti:2003qa}. Their equations of motion are
\begin{equation}\label{ft1}
\gamma^n\,\partial_n\,\Psi_{m_1\cdots
m_{s-\frac{1}{2}}}=(s-\frac{1}{2})\,\partial_{(m_1}\,\chi_{m_2\cdots
m_{s-\frac{1}{2}})}\,,
\end{equation}
\begin{equation}\label{ft2}
\partial^n\,\Psi_{nm_2\cdots
m_{s-\frac{1}{2}}}-(s-\frac{3}{2})\,\partial_{(m_2}\,\lambda_{m_3\cdots
m_{s-\frac{1}{2}})}=\gamma^n\,\partial_{n}\,\chi_{m_2\cdots
m_{s-\frac{1}{2}}}\,,
\end{equation}
\begin{equation}\label{ft3}
\gamma^n\,\partial_n\,\lambda_{m_1\cdots
m_{s-\frac{5}{2}}}=(s-\frac{5}{2})\,\partial^{n}\,\chi_{nm_1\cdots
m_{s-\frac{5}{2}}}\,.
\end{equation}

These equations are invariant under the following unconstrained
gauge transformations of the fields
\begin{equation}\label{gtft1}
\delta\,\Psi_{m_1\cdots
m_{s-\frac{1}{2}}}=(s-\frac{1}{2})\,\partial_{(m_1}\,\xi_{m_2\cdots
m_{s-\frac{1}{2}})}\,,
\end{equation}
\begin{equation}\label{gtft2}
\delta\,\chi_{m_1\cdots
m_{s-\frac{3}{2}}}=\gamma^n\partial_{n}\,\xi_{m_1\cdots
m_{s-\frac{3}{2}}}\,,
\end{equation}
\begin{equation}\label{gtft3}
\delta\,\lambda_{m_1\cdots
m_{s-\frac{5}{2}}}=\partial^n\,\xi_{nm_1\cdots m_{s-\frac{5}{2}}}\,.
\end{equation}

The form of the gauge transformations (\ref{gtft1})--(\ref{gtft3})
prompts us that the fermionic triplet fields are related to the
components of the fermionic higher--spin `vielbein'
$\psi_{n;m_1\ldots m_{s-\frac{3}{2}}}$ as follows
\begin{equation}\label{Psi}
\Psi_{m_1\cdots m_{s-\frac{1}{2}}}=(s-\frac{1}{2})\,\psi_{(m_1;m_2\ldots
m_{s-\frac{1}{2}})}\,,
\end{equation}
\begin{equation}\label{chi}
\chi_{m_1\cdots m_{s-\frac{3}{2}}}=\gamma^{n}\,\psi_{n;m_1\ldots
m_{s-\frac{3}{2}}}\,,
\end{equation}
\begin{equation}\label{lambda}
\lambda_{m_1\cdots m_{s-\frac{5}{2}}}=\eta^{nl}\,\psi_{n;lm_1\ldots
m_{s-\frac{5}{2}}}\,.
\end{equation}
Upon this identification the fermionic triplet field equations of
motion (\ref{ft1})--(\ref{ft3}) follow from eqs. (\ref{ffe}). As
such, upon substituting eqs. (\ref{Psi})--(\ref{lambda}) for
corresponding components of the fermionic frame--like field into the
action (\ref{feract}) and gauge fixing to zero its Stueckelberg
symmetry, one will reduce eq. (\ref{feract}) to the fermionic
triplet action of \cite{Sagnotti:2003qa} in flat space--time.

\section{Frame--like action for fermionic higher--spin fields in
AdS$_D$}\label{AdSftriplets}

 In AdS space the gauge transformations (\ref{fgtr})
(for $t=0$) of the dynamical fermionic field $\psi^\alpha_{a_1\cdots
a_{s-{3\over 2}}}$ are modified as follows \footnote{The form of the
gauge transformations (\ref{fgtr}) of the   higher rank extra fields
(with $t\geq 1$) is more involved. It is not needed for our
consideration, however.}
\begin{equation}\label{fgtrAds}
\delta\psi_{a_1\cdots a_{s-{3\over 2}}}={\mathcal D}\,\xi_{a_1\cdots a_{s-{3\over
2}}}-(s-\frac{3}{2})\,e^b\,\xi_{a_1\cdots a_{s-{3\over 2}},b}\,,
\end{equation}
where following \cite{fronsdal78b} the generalized covariant
differential ${\mathcal D}$ is defined as the sum of the
conventional AdS covariant differential $\nabla$ and the term
${i{\sqrt{-\Lambda}}\over 2}\,e^a\,\gamma_a$, namely,
\begin{equation}\label{mathcalD}
{\mathcal D}=\nabla+{i{\sqrt{-\Lambda}}\over 2}\,e^a\,\gamma_a\,.
\end{equation}
The exterior differential (\ref{mathcalD}) is actually covariant
with respect to the AdS isometry group \mbox{$Spin(2,D-1)$}. It is
defined in such a way that its square vanishes when acting on spinor
differential forms
\begin{equation}\label{D2s}
{\mathcal D}^2\,\chi^\alpha=0
\end{equation}
and it acts as $\nabla^2$ on the tensor differential forms
\begin{equation}\label{D2s1}
{\mathcal D}^2\,T^{a_1\cdots a_t}=\nabla^2\,T_{a_1\cdots
a_t}=-t\,\Lambda\,e^{(a_1}\,e_b\,T^{a_2\cdots a_t)b}\,.
\end{equation}
Thus, in virtue of eq. (\ref{D2s}), ${\mathcal D}^2$ acts on the
tensor--spinors in the same way as on the tensors, \emph{i.e.}
\begin{equation}\label{D2s2}
{\mathcal D}^2\,\psi^{a_1\cdots
a_t}=-t\,\Lambda\,e^{(a_1}\,e_b\,\psi^{a_2\cdots a_t)b}.
\end{equation}
Note also that
\begin{equation}\label{Dgamma}
{\mathcal D}\,\gamma_a=-{i{\sqrt{-\Lambda}}\over
2}\,e^b\,[\gamma_b,\gamma_a]=-i\,\sqrt{-\Lambda}\,e^b\,\gamma_{ba}\,.
\end{equation}
Eqs. (\ref{D2s})--(\ref{Dgamma}) are useful when checking the gauge
invariance of the action for the fermionic higher--spin fields in
AdS under the transformations (\ref{fgtrAds}).

\subsection{Fang-Fronsdal case in AdS}
As in the flat--space, the fermionic dynamical higher--spin field
$\tilde\psi^{a_1\cdots a_{s-{3\over 2}}}$ in AdS is subject to the
gamma--trace condition
\be\label{gtrads}
\gga^c \tilde\psi_{a_1\ldots a_{s-\frac{5}{2}} c}=0\,.
\ee
For this condition to be compatible with the gauge transformations
(\ref{fgtrAds}) the gauge parameters must obey the following
constraints
\bee\label{gtrparametersads}
&\hspace{-50pt}\gga^c \tilde\xi_{a_1\ldots a_{s-\frac{5}{2}}
c}=0\,,\quad (s-\frac{3}{2})\,\gga^c
\tilde\xi_{a_1\ldots a_{s-\frac{5}{2}}
c,b}=i\,\sqrt{-\Lambda}\,\gamma_b{}^c\,\tilde\xi_{a_1\ldots
a_{s-\frac{5}{2}}c}=-{i\,\sqrt{-\Lambda}}\,\tilde\xi_{a_1\ldots
a_{s-\frac{5}{2}}b}&\\
&\hspace{-20pt}
\Longrightarrow \quad \gga^b \tilde\xi_{a_1\ldots
a_{s-\frac{5}{2}} c,b}={i\,\sqrt{-\Lambda}}\,\tilde\xi_{a_1\ldots
a_{s-\frac{5}{2}}c}.&\nonumber
\eee
The action for $\gamma$--traceless $\tilde\psi_{a_1\ldots
a_{s-\frac{3}{2}}}$ in AdS, that is invariant under (\ref{fgtrAds})
with the parameters satisfying (\ref{gtrparametersads}) has the
following form
\bee \label{feractads}\hspace{-10pt} S= i\int_{M^D}e^{a_1} \ldots
e^{{a_{D-3}}}\,\varepsilon_{a_1 \ldots a_{D-3}pqr} \left[
\bar{\tilde\psi}_{d_1\ldots d_{s-{3\over 2}}} \gamma^{pqr} {\mathcal D}\,\tilde\psi^{d_1\ldots
d_{s-{3\over 2}}} -6(s-\frac{3}{2})\, \bar{\tilde\psi}_{d_1\ldots
d_{s-{5\over 2}}}{}^p\gamma^{q}
\,{\mathcal D}\,\tilde\psi^{d_1
\ldots d_{s-{5\over 2}} }{}^r\right.\nonumber\\
\left.+\,{{{3i\,\sqrt{-\Lambda}\,(s-\frac{3}{2})}}\over{D-2}}\,\left(e^r\,
\bar{\tilde\psi}_{d_1\ldots d_{s-{3\over 2}}}\gamma^{pq}\tilde\psi^{d_1\ldots d_{s-{3\over 2}}}
+2(s-\frac{3}{2})\,e^p\,\bar{\tilde\psi}^q{}_{d_1\ldots d_{s-{5\over
2}}}\tilde\psi^{rd_1\ldots d_{s-{5\over
2}}}\right)\right]\,.\hspace{30pt}
 \eee
The last two ``mass--like" terms in (\ref{feractads}) are
proportional to the square root of the cosmological constant (which
is also present in the covariant differential $\mathcal D$
(\ref{mathcalD})). These terms insure the gauge invariance of the
higher--spin system in AdS.

\subsection{Fermionic triplets in AdS}\label{ftads}
Let us now consider the form of the action in AdS space for the
fermionic higher--spin fields $\psi^{a_1\cdots a_{s-{3\over 2}}}$
which are not subject to the gamma--trace condition, \emph{i.e.}
describe fermionic triplets. By now the action and the equations of
motion for the fermionic triplets have been unknown. To demonstrate
that such an action and equations of motion do exist, we first
consider the simplest case of the reducible field of spin
${5\over 2}$.

\subsubsection{Spin--${5\over 2}$ example}
The one--form tensor--spinor field under consideration is the
gamma--traceful field $\psi^a=dx^m\,\psi^a_m$. Its gauge
transformations have the form
\begin{equation}\label{g52}
\delta\psi^a={\mathcal D}\xi^a-e^b\xi^{a,b},
\end{equation}
where the parameter $\xi^a$ is gamma--traceful, while the
antisymmetric parameter $\xi^{a,b}=-\xi^{b,a}$ is required to
satisfy the following relation
\begin{equation}\label{g521}
\gamma_b\,\xi^{a,b}=-i\sqrt{-\Lambda}\,\gamma^{ab}\xi_b
=i\sqrt{-\Lambda}\,(\xi^a-\gamma^a\,\gamma^b\,\xi_b).
\end{equation}
The condition (\ref{g521}), which reduces to the corresponding eq.
(\ref{gtrparametersads}) in the gamma--traceless case, ensures that
the gamma trace of $\psi^a$ transforms as a divergence,\emph{ i.e.}
as a Rarita--Schwinger field of spin 3/2,
\begin{equation}\label{g523}
\delta(\gamma_a\,\psi^a)={\mathcal D}(\gamma_a\,\xi^a).
\end{equation}
The action for the field $\psi^a$ which is invariant under the
transformations (\ref{g52})--(\ref{g523}) has the following form
\bee \label{feractadst52}S= i\int_{M^D}e^{a_1} \ldots
e^{{a_{D-3}}}\,\varepsilon_{a_1 \ldots a_{D-3}bcd} \left[
\bar{\psi}_{f} \gamma^{bcd} {\mathcal D}\,\psi^{f} -6\, \bar{\psi}^b\gamma^{c}
\,{\mathcal D}\,\psi^d
+\,{{{3i\,\sqrt{-\Lambda}\,}}\over{D-2}}\left(\,e^d\,
\bar\psi_{f}\gamma^{bc}\psi^{f}
\right.\right.\nonumber\\
\left.\left.
+2\,e^b\,\bar\psi^c{}\psi^{d}
 +2\,e^d\,
(\bar\psi_{f}\gamma^f)\,\gamma^{b}\psi^{c} -\,e^d\,
(\bar\psi_{f}\gamma^f)\,\gamma^{bc}\,(\gamma_i\psi^{i})\right)\right]\,.
 \eee
One can see that in comparison with the action (\ref{feractads}) for
a single spin--5/2 field, the action (\ref{feractadst52}) contains
two more terms which depend on the gamma--trace of $\psi^{a}$. It
can be shown that by splitting $\psi^{a}$ into the gamma-traceless
and gamma-trace parts
\begin{equation}\label{split}
\psi^{a}=\tilde \psi^a-{1\over D}\gamma^a\,\tilde\psi, \qquad
\gamma_a\,\tilde\psi^{a}= 0, \qquad \tilde\psi=\gamma_a\,\psi^{a}\,,
\end{equation}
the action (\ref{feractadst52}) splits into the direct sum of the
actions for the single spin--5/2 field $\tilde\psi^a$ and the
spin--3/2 field $\tilde\psi$ in a way similar to the bosonic case
(see Subsection
\ref{tripletcase1}). As mentioned above, the spin--1/2 field does not
appear in our construction. The above example is the simplest
fermionic ``triplet" (actually the doublet) of fields in AdS space
\be\label{doublet}
\Psi_{ab}=2\psi_{(b;a)}\,,\qquad \chi_a=\gamma^b\,\psi_{b;a}\,.
\ee
Their gauge transformations are
\be\label{doubletg}
\delta\Psi_{ab}=2{\mathcal D}_{(b}\xi_{a)}\,,\qquad \delta\chi_a=\gamma^b\,{\mathcal D}_{b}\xi_{a}-
i\sqrt{-\Lambda}\,\gamma_{a}{}^b\,\xi_b\,.
\ee

\subsubsection{Generic higher--spin fermion triplets in AdS}
The gamma--traceful one--form tensor--spinor field $\psi^{a_1\cdots
a_{s-{3\over 2}}}$ describing the fermionic triplet in AdS space
undergoes the gauge transformations
\begin{equation}\label{fgtrAdst}
\delta\psi^{a_1\cdots a_{s-{3\over 2}}}={\mathcal D}\,\xi^{a_1\cdots a_{s-{3\over
2}}}-(s-\frac{3}{2})\,e_b\,\xi^{a_1\cdots a_{s-{3\over 2}},b}\,
\end{equation}
with the unconstrained parameter $\xi^{a_1\cdots a_{s-{3\over 2}}}$
and the Stueckelberg parameter $\xi^{a_1\cdots a_{s-{3\over 2}},b}$
satisfying the Young tableau property, $\xi^{(a_1\cdots a_{s-{3\over
2}},b)}=0$, the relaxed traceless condition (as in the case of the
bosonic triplets)
\begin{equation}\label{gtrparametersadstr}
\xi^{a_1\cdots a_{s-{5\over 2}}c,b}\,\eta_{bc}=0
\end{equation}
and the following relation
\begin{eqnarray}\label{gtrparametersadst}
\gga_b \xi^{a_1\ldots
a_{s-\frac{3}{2}},b}=-\,{i\,\sqrt{-\Lambda}}\gamma^{(a_1}{}_{b}\,\xi^{a_2\ldots
a_{s-\frac{3}{2}})b}\,.
\quad
\end{eqnarray}
Eq. (\ref{gtrparametersadst}) reduces to  (\ref{gtrparametersads})
if the parameter $\xi^{a_1\cdots a_{s-{3\over 2}}}$ was
gamma--traceless and ensures that the gamma--trace of
$\psi^{a_1\cdots a_{s-{3\over 2}}}$ transforms as a spin--$(s-1)$
field, \emph{i.e.} similarly to (\ref{fgtrAdst}) with
$s\,\rightarrow\, s-1$.

The action, that is invariant under the transformations
(\ref{fgtrAdst})--(\ref{gtrparametersadst}), has the following form
\bee \label{feractadst}
&S= i\int_{M^D}e^{a_1} \ldots e^{{a_{D-3}}}\,\varepsilon_{a_1 \ldots
a_{D-3}abc} \left[
\bar{\psi}_{d_1\ldots d_{s-{3\over 2}}} \gamma^{abc} {\mathcal D}\,\psi^{d_1\ldots
d_{s-{3\over 2}}} -6(s-\frac{3}{2})\, \bar{\psi}_{d_1\ldots
d_{s-{5\over 2}}}{}^a\gamma^{b}
\,{\mathcal D}\,\psi^{d_1
\ldots d_{s-{5\over 2}} }{}^c\right. 
\nonumber\\
&\left.\right.\nonumber\\
&\left.+\,{{{3i\,\sqrt{-\Lambda}\,(s-\frac{3}{2})}}\over{D-2}}\,\left(e^c\,
\bar\psi_{d_1\ldots d_{s-{3\over 2}}}\gamma^{ab}\psi^{d_1\ldots d_{s-{3\over 2}}}
+2(s-\frac{3}{2})\,e^a\,\bar\psi^b{}_{d_1\ldots d_{s-{5\over
2}}}\psi^{cd_1\ldots d_{s-{5\over 2}}}\right)
\right.\nonumber\\
&\left.
\right.\\
&\left.+
{{{3i\,\sqrt{-\Lambda}}\,(s-\frac{3}{2})}\over{D-2}}\,\left(2\,e^c\,
(\bar\psi_{d_1\ldots d_{s-{5\over
2}}f}\gamma^f)\,\gamma^{a}\psi^{bd_1\ldots d_{s-{5\over 2}}}
-\,e^c\, (\bar\psi_{d_1\ldots d_{s-{5\over
2}}f}\gamma^f)\,\gamma^{ab}\,(\gamma_i\psi^{id_1\ldots d_{s-{5\over
2}}})\right)
\right.\nonumber\\
&\left.\right.\nonumber\\
&\left.-{{{6i\,\sqrt{-\Lambda}}\,(s-\frac{3}{2})(s-\frac{5}{2})}\over{D-2}}\,\,e^a\,(\bar\psi^{bi}{}_{d_1\ldots
d_{s-{7\over 2}}}\gamma_{i})\,(\gamma_f\psi^{cfd_1\ldots
d_{s-{7\over 2}}})
\right]\,.
\nonumber
 \eee
It has one more (the last) term in comparison with the action
(\ref{feractadst52}) for the $s=\frac{5}{2}$ triplet.

The AdS analogues of the flat--space fermionic triplet fields of
\cite{Francia:2002pt,Sagnotti:2003qa} are extracted from
$\psi_{a_1\cdots a_{s-{3\over 2}}}=e^b\psi_{b;a_1\cdots a_{s-{3\over
2}}}$ analogously to (\ref{Psi})--(\ref{lambda})
\begin{equation}\label{Psiads}
\Psi_{a_1\cdots a_{s-\frac{1}{2}}}=(s-\frac{1}{2})\,\psi_{(a_1;a_2\ldots
a_{s-\frac{1}{2}})}\,,
\end{equation}
\begin{equation}\label{chiads}
\chi_{a_1\cdots a_{s-\frac{3}{2}}}=\gamma^{b}\,\psi_{b;a_1\ldots
a_{s-\frac{3}{2}}}\,,
\end{equation}
\begin{equation}\label{lambdaads}
\lambda_{a_1\cdots a_{s-\frac{5}{2}}}=\eta^{bc}\,\psi_{b;ca_1\ldots
a_{s-\frac{5}{2}}}\,.
\end{equation}

Their gauge transformations are easily obtained from eqs.
(\ref{fgtrAdst})--(\ref{gtrparametersadst})
\begin{equation}\label{gtft1ads}
\delta\,\Psi_{a_1\cdots
a_{s-\frac{1}{2}}}=(s-\frac{1}{2})\,{\mathcal
D}_{(a_1}\,\xi_{a_2\cdots a_{s-\frac{1}{2}})}\,,
\end{equation}
\begin{equation}\label{gtft2ads}
\delta\,\chi_{a_1\cdots
a_{s-\frac{3}{2}}}=\gamma^b{\mathcal D}_{b}\,\xi_{a_1\cdots
a_{s-\frac{3}{2}}}-(s-{3\over
2})\,{i\,\sqrt{-\Lambda}}\gamma_{(a_1}{}^{b}\,\xi_{a_2\ldots
a_{s-\frac{3}{2}})b}\,,
\end{equation}
\begin{equation}\label{gtft3ads}
\delta\,\lambda_{a_1\cdots
a_{s-\frac{5}{2}}}={\mathcal D}^b\,\xi_{ba_1\cdots
a_{s-\frac{5}{2}}}\equiv \nabla^b\,\xi_{ba_1\cdots
a_{s-\frac{5}{2}}}+\frac{i\sqrt{-\Lambda}}{2}\,\gamma^b\,\,\xi_{ba_1\cdots
a_{s-\frac{5}{2}}}\,,
\end{equation}
and the equations of motion, which generalize to AdS space eqs.
(\ref{ft1})--(\ref{ft3}), follow from the action (\ref{feractadst}).

Note that the gauge transformations (\ref{gtft2ads}) and
(\ref{gtft3ads}) of the fields $\chi$ and $\lambda$ of the fermionic
triplet in AdS differ from those given in
\cite{Buchbinder:2007ak} by terms proportional to
the gamma--trace of the gauge parameter $\gamma^{b}\,\xi_{ba_1\cdots
a_{s-\frac{5}{2}}}$. We assume that this is a reason which have not
allowed previous authors to obtain the Lagrangian description of the
fermionic triplets in AdS.

To recapitulate, in the frame--like formulation the fermionic
triplet is described by the unconstrained fermionic higher--spin
vielbein $dx^m\,\psi_{m;a_1\cdots a_{s-\frac{3}{2}}}$ subject to the
gauge transformations (\ref{fgtrAdst}) with the Stueckelberg
parameters $\xi_{a_1\cdots a_{s-\frac{3}{2}},b}$ satisfying the
relaxed (gamma)--trace constraints (\ref{gtrparametersadstr}) and
(\ref{gtrparametersadst}). Upon eliminating  the Stueckelberg
degrees of freedom and splitting the components of the spinor-tensor
$\psi_{b;a_1\cdots a_{s-{3\over 2}}}$ into its triplet constituents
(\ref{Psiads})--(\ref{lambdaads}) one can reduce the action
(\ref{feractadst})  to an action which describes the fermionic
triplets in AdS in the metric--like formulation.

\section{Relation to unconstrained formulations of irreducible
higher--spin fields}

Let us now demonstrate how the triplet systems discussed in the
previous sections can be reduced to corresponding irreducible fields
of spin $s$ without imposing conventional (gamma)--trace constraints
on the fields and gauge parameters. The consideration below applies
both to the flat space--time and to the AdS background.

We observe that the action $S^{irr}$ for the irreducible spin $s$
system results from the action $S^{red}$ for the reducible (triplet)
system by adding the Lagrange multiplier term
\be\label{sirr}
S^{irr} = S^{red}+
\int\,l_{a_1\cdots a_{s-3}}\,e^{a_1\cdots a_{s-3} c}{}_c\,,
\ee
where $l^{a_1\cdots a_{s-3}}=dx^{m_1}\,\cdots
dx^{m_{D-1}}\,l_{m_1\cdots m_{D-1}}^{a_1\cdots a_{s-3}}(x)$ is a
(frame--like) differential $(D-1)$--form Lagrange multiplier, which
is assumed to be gauge invariant. The Lagrange multiplier term  in
(\ref{sirr}) is not invariant under the full relaxed gauge symmetry
transformations, but only under those with traceless parameters. The
full relaxed gauge invariance can be restored, however, by making
the following substitution in the action (\ref{sirr})
\begin{eqnarray}\label{subs}
e^{a_1\cdots a_{s-3}c}{}_c
\longrightarrow e^{a_1\cdots a_{s-3}c}{}_c -
{\nabla}\,\alpha^{a_1\cdots a_{s-3}}\,+(s-1)\,e^b\,\beta^{a_1\cdots
a_{s-3},}{}_{b}\,,\hspace{70pt}\\
\omega_c{}^{ca_1\cdots a_{s-3},b} \longrightarrow \omega_c{}^{ca_1\cdots a_{s-3},b}
-\nabla\,\beta^{a_1\cdots a_{s-3},b}+\Lambda\,(e^b\,\alpha^{a_1
\ldots a_{s-3}}-e^{(a_1}\,\alpha^{a_2\ldots a_{s-3})b})\,,
\nonumber
\end{eqnarray}
where $\alpha^{a_1\cdots a_{s-3}}$ and $\beta^{a_1\cdots a_{s-3},m}$
are zero--form Stueckelberg fields (\emph{i.e.} compensators) (the
latter having the symmetry of the Young tableau $Y(s-3,1)$). To make
the final action compatible with the transformation rules
(\ref{gs1}) and (\ref{trmn}) in flat space or (\ref{gsads11}) and
(\ref{rtads1}) in AdS, the compensator fields are endowed with the
following transformation laws
\be\label{comp1}
\delta\,\alpha^{a_1\cdots
a_{s-3}}=\xi^{a_1\cdots a_{s-3}b}{}_b\,,
\ee
\be\label{comp2}
\delta\,\beta^{a_1\cdots a_{s-3},b}=\xi^{a_1\cdots
a_{s-3}}{}_c{}^{c,b}\,.
\ee

Let us stress that since the $S^{red}$ is invariant under the full
relaxed higher-spin gauge transformations, it obviously remains
intact under the substitution (\ref{subs}), \emph{i.e.} does not
contain the compensator fields. Thus, the resulting compensator
action for the irreducible field of spin $s$ is
\be
\label{puract}
S^{irr} = S^{red}+
\int\,l_{a_1\cdots a_{s-3}}(
 e_{b;}{}^{a_1\cdots a_{s-3}c}{}_c -
{\nabla}_b\,\alpha^{a_1\cdots a_{s-3}}\,+(s-1)\,\beta^{a_1\cdots
a_{s-3},}{}_{b})\,.
\ee
By construction the action (\ref{puract}) is of the first order in
derivatives and is invariant under the relaxed gauge transformations
similar to those of the reducible triplet system.

The compensator field $\alpha^{a_1\cdots a_{s-3}}$ is nothing but
the one considered for the spin--3 case already by Schwinger
\cite{Schwinger:1970xc} and for arbitrary spin in
\cite{Francia:2002aa,Francia:2002pt,Francia:2005bu,Sagnotti:2003qa,Buchbinder:2007ak,Francia:2007ee},
while the field $\beta^{a_1\cdots a_{s-3},b}$ is a new one, it
``compensates" the Stueckelberg gauge transformations of the trace
of the higher--spin vielbein. (If we imposed the gauge in which the
compensator fields are zero, the system would have reduced to the
conventional Fronsdal case.) The Lagrange multiplier $l_{a_1\cdots
a_{s-3}}$ is the frame--like
 counterpart of the
gauge--invariant Lagrange multipliers of the unconstrained
formulation of
\cite{Buchbinder:2007ak}. More precisely, up to the coefficients,
 the Lagrange multipliers
$\lambda^{a_1\cdots a_{s-2}}$ and $\lambda^{a_1\cdots a_{s-4}}$ of
\cite{Buchbinder:2007ak} are
\be\label{lm1}
\lambda^{a_1\cdots a_{s-2}}=\varepsilon^{b_1\cdots b_{D-1}(a_1}\,l_{b_1\cdots b_{D-1}}^{a_2\cdots
a_{s-2})}\,,\qquad
\lambda^{a_1\cdots a_{s-4}}=\varepsilon^{b_1\cdots b_{D-1}}{}_{b}\,l_{b_1\cdots b_{D-1}}^{a_1\cdots
a_{s-4}b}\,.
\ee

The constraint on the higher--spin vielbein that follows from  the
action (\ref{puract})
\begin{equation}\label{pureg}
e_{b;}{}^{a_1\cdots a_{s-3}c}{}_c={\nabla}_b\,\alpha^{a_1\cdots
a_{s-3}}\,-(s-1)\,\beta^{a_1\cdots a_{s-3},}{}_{b}\,
\end{equation}
implies that the trace of the higher--spin vielbein is pure gauge,
\emph{i.e.} it can be set to zero by gauge fixing the compensators
to zero using (\ref{comp1}) and (\ref{comp2}).

Let us now show that eq. (\ref{pureg}) reduces to corresponding
equations of \cite{Buchbinder:2007ak}. To this end, we symmetrize
all the indices of (\ref{pureg}). In view of the Young symmetry
property $\beta^{(a_1\cdots a_{s-3},b)}\equiv 0$, the result is
\begin{equation}\label{pureg1}
e^{(b;a_1\cdots a_{s-3})c}{}_c={\nabla}^{(b}\,\alpha^{a_1\cdots
a_{s-3})}\,.
\end{equation}
Using the triplet field redefinition (\ref{ephiD}) (or
(\ref{ephiDads})) we can rewrite eq. (\ref{pureg1}) in the form
\begin{equation}\label{puregDF}
D^{a_1\cdots a_{s-2}}-{1\over 2}\,\Phi^{a_1\cdots
a_{s-2}c}{}_c=-\frac{(s-2)}{2}\,{\nabla}^{(a_1}\,\alpha^{a_2\cdots
a_{s-2})}\,.
\end{equation}

Let us now contract  the index $b$ in (\ref{pureg}) with one of the
indices $a_i$. We get another condition that
\begin{equation}\label{pureg2}
e_{b;}{}^{a_1\cdots a_{s-4}bc}{}_c={\nabla}_b\,\alpha^{a_1\cdots
a_{s-4}b}\,,
\end{equation}
or, in view of eqs. (\ref{ephiD}) and (\ref{ephiDads}),
\begin{equation}\label{puregD}
D^{a_1\cdots a_{s-4}b}{}_b={\nabla}_b\,\alpha^{a_1\cdots
a_{s-4}b}\,.
\end{equation}
Up to a field rescaling, eqs. (\ref{puregDF}) and (\ref{puregD}) are
the ones which appeared in minimal unconstrained formulations
\cite{Sagnotti:2003qa,Buchbinder:2007ak,Francia:2007ee} of the
irreducible higher--spin fields (for non--minimal formulations see
\emph{e.g.} \cite{Pashnev:1997rm,Burdik:2001hj,Buchbinder:2001bs,Barnich:2004cr}).

The fermionic case can be considered in a similar fashion. The
additional requirement that the gamma--trace of the higher--spin
field $\psi_{m;\,a_1\ldots a_{s-{3\over 2}}}$ is a pure gauge,
\emph{i.e.} the constraint that should be added to the triplet
action (\ref{feract}) or (\ref{feractadst}) with the Lagrange
multiplier to reduce the fermionic triplet to the irreducible field
of spin $s$ is
\be
\label{fgtrg}
\gamma^{a_{s-{3\over 2}}}\, \psi_{m;\,a_1\ldots a_{s-{5\over 2}}a_{s-{3\over 2}}}=
{\mathcal D}_m\,\alpha_{a_1\ldots a_{s-{5\over
2}}}-(s-\frac{3}{2})\,
\beta_{a_1\ldots a_{s-{5\over 2}},\,m}\,
\ee
(with $\alpha_{a_1\ldots a_{s-{5\over 2}}}$ and $\beta_{a_1\ldots
a_{s-{5\over 2}},\,m}$ being fermionic compensators). By construction it
is of first order in derivatives.

We have thus,  obtained the frame--like version of the unconstrained
Lagrangian formulations of the irreducible higher--spin fields
considered in \cite{Buchbinder:2007ak,Francia:2007ee}.

\section{AdS covariant formalism for bosonic HS fields}
\label{ADSC}
\setcounter{equation}0

The description of the higher--spin fields in AdS space considered
in the previous sections was not manifestly invariant under the
higher--spin gauge transformations. Also it was  manifestly
invariant only under the Lorentz subgroup $O(1,D-1)$ of the full AdS
isometry group $O(2,D-1)$. To make both the AdS isometry and
higher--spin gauge symmetries manifest it is convenient to use the
formalism a la MacDowell, Mansouri, Stelle and West
\cite{MM,SW} (MMSW). We shall introduce only basic ingredients of
this formulation which are required for our purposes and refer the
reader to \cite{5d,bciv} for further details.

\subsection{Basic definitions}
The $AdS_D$ space is described by the vielbein $e^a=dx^m\,e^a_m$ and
the connection $\omega^{ab}=dx^m\,\omega^{ab}_m$ which satisfy the
zero torsion and constant curvature conditions (\ref{adst}) and
(\ref{adsR}). To make the $O(2,D-1)$ AdS isometry symmetry manifest
we unify $e^a$ and $\omega^{ab}$ into a connection $\Omega^{AB}$
valued in the algebra $o(2,D-1)$
\begin{equation}\label{OmegaAdS}
\Omega^{AB}:=(\omega^{ab},\sqrt{-\Lambda}\,e^a) \q
i.e. \qquad e^{a}={1\over\sqrt{-\Lambda}}\,\Omega^{a0'}\,,
\end{equation}
where the capital Latin indices $A,B,\cdots=(0',a) =
(0',0,1,\cdots\,,D-1)$ correspond to the vector representation of
$o(2,D-1)$ acting in a $D+1$ dimensional vector space with the
invariant metric $\eta_{AB}=(+,+,-,\cdots\,,-)$, and the index $0'$
denotes the second time--like direction in this space. Recall that
the cosmological constant $\Lambda$ is negative in the AdS case.

The connection $\Omega^{AB}$ that satisfies the zero curvature
equation
\begin{equation}\label{calR}
{\mathcal R}:=d\Omega+\Omega^2=0\,.
\end{equation}
promotes the rigid isometry symmetry $O(2,D-1)$ to the local one.

By construction eq. (\ref{calR}) is equivalent to the relations
(\ref{adst}) and (\ref{adsR}) satisfied by the AdS torsion and
curvature.

Because of the zero curvature condition (\ref{calR}) it is
convenient to work with the exterior covariant derivative associated
with the connection $\Omega$, that squares to zero in virtue of
(\ref{calR})
\begin{equation}\label{calD}
{\mathcal D}=d+\Omega\,,\qquad {\mathcal D}{\mathcal D}={\mathcal
R}=0\,.
\end{equation}
Note that the exterior covariant derivative (\ref{mathcalD}) which
we used to describe the fermionic fields in AdS in Section
\ref{AdSftriplets} is nothing but the covariant derivative
 (\ref{calD}) in the spinor representation of $Spin(2,D-1)$ with the
spin connection ${1\over{2}} \Omega^{AB}\,{\Gamma_A \Gamma_B}$.  The
matrices $\Gamma_A$ are Dirac matrices corresponding to the group
$Spin(2,D-1)$. The Dirac matrices $\gamma^a$ corresponding to
$Spin(1,D-1)$ are related to $\Gamma^A$ as follows
$$
\gamma^a=i\Gamma^a\,\Gamma^{0'}\,.
$$

Another ingredient of the MMSW formulation is the so called
compensator vector field $V^A(x)$ satisfying the normalization
condition
\be V^A V^B \eta_{AB} =-\frac{1}{\Lambda}. \label{norm}
\ee
The extension of the symmetry from the local Lorentz group
$O(1,D-1)$ to $O(2,D-1)$ brings about $D$ more local symmetry
parameters, which can be regarded as coordinates for the coset space
$O(2,D-1)/O(1,D-1)$. The role of the field $V^A(x)$ is similar to
that of the Goldstone fields. It compensates the action of these
additional local symmetries and thus maintains intact the number of
the physical degrees of freedom of the model. Using local $O(2,D-1)$
transformations of $V^A$ one can choose the gauge
\begin{equation}\label{vgauge}
V^A={1\over{\sqrt{-\Lambda}}}\delta^A_{0'}
\end{equation}
which breaks the local symmetry $O(2,D-1)$ down to $O(1,D-1)$. The
one--form
\be\label{cv}
E^A={\mathcal D}\,V^A
\ee
is the $O(2,D-1)$--covariant vielbein. It reduces to $e^a$ in the
gauge (\ref{vgauge}).

In the AdS-covariant formulation, the dynamics of massless
higher--spin fields is described \cite{5d} by the generalized
connection one--form
\begin{equation}\label{Om}
\Omega^{A_1 \cdots\, A_{s-1},\, B_1 \cdots\, B_{s-1}}(x)
=dx^m\,\Omega_m^{A_1 \cdots\, A_{s-1},\, B_1 \cdots\, B_{s-1}}
\qquad (A,B= 0',0,1\cdots\,,D-1)\,,
\end{equation}
that takes values in the $O(2,D-1)$--module described by the
two--row rectangular Young tableau
\begin{picture}(60,15)(-5,2) \multiframe(0,7.5)(13.5,0){1}(50,7){}
\multiframe(0,0)(13.5,0){1}(50,7){}\put(53,0){}
\end{picture}
of length $s-1$ and, hence, satisfies the symmetry conditions
\be \Omega_m^{A_1\cdots\, A_{s-1}\, ,\, B_1\cdots\,
B_{s-1}}=\Omega_m^{(A_1\cdots\, A_{s-1})\, ,\, B_1\cdots\,
B_{s-1}}=\Omega_m^{A_1\cdots\, A_{s-1}\, ,\, (B_1\cdots\,
B_{s-1})}\,, \label{adsy1}\ee
\be\Omega_m^{(A_1\cdots\, A_{s-1}\, ,\, A_s)\,B_2\cdots\,
B_{s-1}}=0\,. \label{adsy2}\ee
As a consequence of eqs.~(\ref{adsy1}) and (\ref{adsy2}), the
higher--spin connection is (anti)symmetric with respect to the
interchange of the two groups of  symmetrized indices:
\be \Omega_m^{A_1\cdots\, A_{s-1},B_1\cdots\,
B_{s-1}}=(-1)^{s-1}\,\Omega_m^{B_1\cdots\, B_{s-1},A_1\cdots\,
A_{s-1}}\,. \label{AB}\ee

The linearized higher--spin curvature associated with this
connection is
\bqn\label{R1}
 {\mathcal R}^{A_1
\cdots A_{s-1}, B_1 \cdots B_{s-1}}=
{\mathcal D} \,\Omega^{A_1 \cdots A_{s-1}, B_1 \cdots B_{s-1}}\hspace{250pt} \nonumber \\
\\
= d \Omega^{A_1 \cdots A_{s-1}, B_1 \cdots B_{s-1}}+
(s-1)\,\Omega_{~~~C}^{~(A_1}\,
\Omega^{A_2 \cdots  A_{s-1})C, \,B_1 \cdots B_{s-1}} 
- (s-1)\, \Omega^{A_1 \cdots  A_{s-1},
\,C (B_2
\cdots B_{s-1}} \,\Omega_{~~~C}^{~B_1)} \;, \nonumber 
\eqn
where $\Omega^{AB} $ is the AdS background $O(2,D-1)$ spin
connection (\ref{OmegaAdS}).

The higher--spin curvature is invariant under the local
transformations with parameters $\xi(x)$, that have the same
symmetry properties as the higher--spin connection
\begin{equation}\label{gt}
\delta \Omega^{A_1\cdots A_{s-1},C_1\cdots C_{s-1}}={\mathcal
D}\,\xi^{A_1\cdots A_{s-1},C_1\cdots C_{s-1}}\,.
\end{equation}

The irreducible Lorentz components of the connection $\Omega$
contain the higher--spin vielbein and Lorentz connection analogous
to those in flat space--time (\ref{viel}) and (\ref{connect}) as
well as all extra connections. They result from (\ref{Om}) by
projecting $\Omega$ along the compensator $V^A$. For instance,
\begin{equation}\label{Vcontruction1}
 \Omega^{A_1\cdots A_{s-1},C_1}:= \Omega^{A_1\cdots
A_{s-1},C_1\cdots C_{s-1}}V_{C_2}\cdots V_{C_{s-1}},
\end{equation}
contains  the higher--spin vielbein
\begin{equation}\label{Vcontruction}
 E^{A_1\cdots A_{s-1}}:=
\Omega^{A_1\cdots A_{s-1},C_1\cdots C_{s-1}} V_{C_1}\cdots
V_{C_{s-1}}\,,
\end{equation}
as the most $V$-longitudinal components of $\Omega$ and the
higher--spin Lorentz connection as next to the most
 $V$-longitudinal components of $\Omega$.
In the gauge (\ref{vgauge}) they are
\begin{eqnarray}\label{Vcontruction2}
 e^{a_1\cdots a_{s-1}}&\equiv&E^{a_1\cdots a_{s-1}}:=
{1\over{(-\Lambda)^{{s-1}\over 2}}}\,\Omega^{a_1\cdots
a_{s-1},0'\cdots
0'}\,,\\
\omega^{a_1\cdots a_{s-1},b}&\equiv&\Omega^{a_1\cdots a_{s-1},b}:=
{1\over{(-\Lambda)^{{s-2}\over 2}}}\,\Omega^{a_1\cdots
a_{s-1},b0'\cdots 0'}.\nonumber
\end{eqnarray}
The gauge transformations of these fields, which follow from
(\ref{gt}), are those given in eqs. (\ref{gsads11}) and
(\ref{gsads12}) but with \emph{traceful} gauge parameters since we
have not imposed the trace constraints on the higher--spin
connection (\ref{Om}) yet.

The other $O(1,D-1)$ tensor fields contained in $\Omega^{A_1\cdots
A_{s-1},B_1\cdots B_{s-1}}$, \emph{i.e.} $\omega^{a_1\cdots
a_{s-1},b_1,\cdots b_t}$ (with $2\geq t \leq s-1$) are the extra
fields which play an important role in interacting higher--spin
systems as shown in \cite{fvnp}.

If the connection $\Omega^{A_1\cdots A_{s-1},B_1\cdots B_{s-1}}$ and
the parameters $\xi^{A_1\cdots A_{s-1},B_1\cdots B_{s-1}}$ are
traceless in the indices $A$ and $B$, they describe a single bosonic
higher--spin field in $AdS _D$ \cite{5d}. The corresponding
$O(1,D-1)$--covariant higher--spin vielbein, connections and gauge
parameters satisfy the trace constraints discussed in Subsection
\ref{adsbf}.

\subsection{Generating functions}
In the previous subsection we have introduced main ingredients of
the AdS covariant description of higher--spin fields characterized
by a definite value of $s$. It is however convenient, and actually
indispensable when constructing higher--spin interactions, to work
simultaneously with the infinite set of spins $s=0,1,\cdots
,\infty$. To this end the formalism of generating functions is most
appropriate.

The space of traceful rectangular two--row Young tableaux of the
algebra $gl(D+1)$ can be conveniently described as the $sp(2)$
invariant subspace $\mathcal V$ of the space of polynomials $f(Y)$
of the variables $Y^A_i$ ($i=1,2$, $A=0,0',1,\ldots D-1$) such that
\be \label{poly} f(Y) =
\sum_{n=0}^\infty f_{A_1\ldots A_n\,,B_1 \ldots B_n} Y^{A_1}_1
\ldots Y^{A_n}_1 Y^{B_1}_2 \ldots Y^{B_n}_2\,,
\ee
\be \label{sp2}
f(Y)\in {\mathcal V} \,:\qquad ( T_{i}{}^j -\frac{1}{2} \delta^i_j
T_{k}{}^k )f(Y) = 0\q T_{i}{}^j = Y^A_i \frac{\partial}{\partial
Y^A_j}\,.
\ee
Note that this system was recently described in analogous fashion in
\cite{alkal}.

Since the conditions (\ref{sp2}) are first order differential
equations,  $\mathcal V$ is in fact the algebra with the pointwise
product law in the $Y$-space, i.e., given two solutions $f_1 (Y)$
and $f_2 (Y)$ of (\ref{sp2}), $f_1 (Y) f_2 (Y)$ also solves the same
condition. Note that the condition (\ref{sp2}) requires in
particular that
\be\label{sp21}
\left (Y^A_1 \frac{\partial}{\partial Y^A_1} - Y^A_2
\frac{\partial}{\partial Y^A_2}\right ) f(Y)=0\,,
\ee
\emph{i.e.} any polynomial in ${\mathcal V}$ contains equal number of $Y_1^A$ and
$Y_2^B$ as in (\ref{poly}). The coefficients $f_{A_1\ldots A_n\,,B_1
\ldots B_n}$  carry various $gl(D+1)$--modules described by two-row
rectangular Young tableaux (see also \cite{alkal}). Note that, for a
homogeneous polynomial of degree $2p$ (equivalently, for a
rectangular two-row Young tableau of length $p$), the condition
(\ref{sp2}) can be rewritten in the form
\be
\label{fc}
 T_{i}{}^j f(Y) = {p}\, \delta^i_j f(Y)\,.
\ee

A useful viewpoint is that the space ${\mathcal V}$ is spanned by
various functions of the elementary $sp(2)$ invariant combinations
$T^{AB}= Y^A_i Y^{Bi}$ where the $sp(2)$ indices are raised and
lowered by the $sp(2)$ invariant symplectic form
\be
\label{symind}
A^i = \gvep^{ij}A_j\q A_i = A^j \gvep_{ji}\,,
\ee
\ie all
$sp(2)$ indices are contracted among themselves. Clearly, the
functions of this class form an algebra.

In terms of the generating functions (\ref{poly})--(\ref{sp2}) the
higher--spin curvatures (\ref{R1}) and gauge transformations
(\ref{gt}) take the following manifestly gauge invariant form
\be\label{RY}
{\mathcal R} ={\mathcal D} \Omega (Y)\,,
\ee
\be\label{xiY}
\delta \Omega (Y)={\mathcal D} \xi(Y)\,,
\ee
where
\be
{\mathcal D} = d +\Omega^{AB} Y_{Ai} \frac{\partial}{\partial
Y^B_i}\,.
\ee

Irreducible  two-row rectangular $o(2,D-1)$--modules are described
by polynomials with traceless coefficients
 $f_{A_1\ldots A_n\,,B_1 \ldots B_n}$, that, in addition to
(\ref{sp2}), satisfy the tracelessness conditions \be \label{tr}
\frac{\partial^2}{\partial Y^A_i \partial Y_{Aj}} f(Y) =0\,.
\ee

The class of functions with the relaxed traceless conditions which
describe the AdS triplet system can be defined as the space
$T\subset \mathcal V$ spanned by various polynomials of the form
(see also \cite{alkal})
 \be
\label{h} h(Y) = \sum_{p=0}^\infty (t(Y))^p h_p (Y)\,,
\ee
where
\be
 \label{t} t(Y) =
 \eta_{ij}(Y) \eta^{ij}(Y)=2\det|\eta_{ij}(Y)| \q \eta_{ij}(Y)= Y^A_i Y_{Aj}\q
 \eta^{ij}(Y) =
  \epsilon^{il}\epsilon^{jk}\eta_{lk}(Y)
\ee%
and $h_p (Y)$ satisfy the conditions (\ref{sp2}) and (\ref{tr})
\be
\label{trh} \frac{\partial^2}{\partial Y^A_i \partial Y_{Aj}} h_p(Y)
=0\,.
\ee
Since,  $t(Y)$ (\ref{t}) is $sp(2)$ invariant, any element (\ref{h})
belongs to the space ${\mathcal V}$ of two-row rectangular Young
tableaux. Although $h(Y)$ is not traceless, its $o(2,D-1)$
irreducible components $h_p (Y)$ all describe two--row rectangular
traceless tensors.

Alternatively, the subspace $T\subset \mathcal V$ can  be described
without explicit reference to the expansion (\ref{h}) as the space
of functions that satisfy the relaxed traceless condition
\be
\label{trc}
t(Y)\frac{\partial^2}{\partial Y^A_i \partial Y_{Aj}}h(Y)  =
\eta^{ij}(Y)
\eta_{kl}(Y) \frac{\partial^2}{\partial Y^C_k \partial Y_{Cl}} h(Y)\,.
\ee
The key observation leading to this condition is that the result  of
action of $\frac{\partial^2}{\partial Y^A_i \partial Y_{Aj}}$ on any
function of the form (\ref{h}) is proportional to $\eta^{ij}(Y)$. It
is then elementary to see that in this case (\ref{trc}) is true.

Thus, the condition (\ref{trc}) singles out only (and all)
rectangular traceless two--row Young tableaux from the generic
traceful two-row Young tableau. Correspondingly,  if we consider a
one-form connection $\Omega(Y)$ that takes values in $T$ and is a
degree $2(s-1)$ polynomial in $Y$
\begin{equation}\label{OmY}
\Omega(Y) = \sum_{p=0}^{[{{s-2}\over 2}]} \,(t(Y))^p \, \Omega_p (Y)\,,
\end{equation}
the traceless components $\Omega_p(Y)$ of this connection correspond
to the set of fields of spins $s, s-2, s-4,\ldots, 3$ or 2. Thus,
such an $\Omega(Y)$ describes a spin--$s$ triplet system (modulo the
scalar and vector fields, as discussed in Section \ref{flmAdS}).

One can easily check that the components of $\Omega(Y)$ (\ref{OmY}),
which form the $2(s-1)$ tensor (\ref{Om})--(\ref{AB}), are related
to the Lorentz covariant components of previous sections via
projection along the compensator field $V^A$. Namely, the Lorentz
irreducible components are singled out by the condition
\begin{equation}\label{VY}
\Omega^{(s-1,t)}{(Y)}=\Pi\left( V^{A_1}\,{\partial\over{\partial Y_2^{A_1}}}
\cdots V^{A_1}\,{\partial\over{\partial
Y_2^{A_{s-1-t}}}}\,\Omega(Y)\right )\,,
\end{equation}
where $\Pi$ is the projector to the $V$--transversal part of $Y^A_i$
\be
\Pi (f(Y)) = f(\Pi (Y))\q \Pi(Y^A_i) = Y^A_i +\Lambda V^A V_B Y^B_i\,.
\ee
The higher--spin vielbein is the most $V$--longitudinal component of
$\Omega$ with $t=0$ in (\ref{VY}). The higher--spin auxiliary
Lorentz--like connection has $t=1$ while extra higher--spin
connections have $t>1$. In the gauge
$V^A=\frac{1}{\sqrt{-\Lambda}}\,\delta^A_{0'}$ the resulting
higher--spin vielbein and connection (\ref{Vcontruction2}) (and
corresponding gauge parameters) have the trace properties of the
triplet system considered in Section \ref{flmAdS}, \ie the
higher--spin vielbein is traceful and the higher--spin connection
satisfies the relaxed traceless condition (\ref{trads}). It can be
also verified that the extra field
$$
\omega^{a_1\cdots
\a_{s-1},b_1b_2} = {1\over{(-\Lambda)^{{s-3}\over 2}}}\,
\Omega^{a_1\cdots \a_{s-1},b_1b_2 0'\cdots 0'}
$$
with  two $o(1,D-1)$ indices in the second row (and the
corresponding Stueckelberg gauge parameter) satisfy the trace
conditions (\ref{xibd})--(\ref{xibd2}). These are consistency checks
of the relation of the AdS $o(2,D-1)$--covariant triplet
construction under consideration with the $o(1,D-1)$--covariant
description of the bosonic triplets of Section
\ref{flmAdS}.

 Let us note that beyond the space $T$,
generic traceful rectangular two-row Young tableaux decompose into a
set of irreducible $o(2,D-1)$ tensors that are not necessarily
described by rectangular two-row Young tableaux (cf \cite{alkal}).
Since one-form connections valued in non-rectangular Young tableaux
describe \cite{SV} so-called partially massless fields \cite{pm}
which correspond to non-unitary representations of the $AdS_D$
algebra $o(2,D-1)$, it is important that these are ruled out of a
quantum-mechanically consistent theory. In this respect, the
relaxation of the tracelessness condition (\ref{tr}) to the relaxed
(triplet) condition (\ref{trc}) is probably the maximal one within
the class of fields that still correspond to a set of unitary
massless fields described by the connections  that take values in
two-row rectangular traceless Young tableaux of $o(2,D-1)$.

\subsection{Action}
 To formulate a manifestly gauge and $o(2,D-1)$--invariant action for
the relaxed system we shall look for it in the form
 bilinear in the manifestly gauge invariant higher--spin
 curvatures (\ref{RY}).

It is convenient to use the version of the formalism of generating
functions introduced in the previous subsection as proposed in
\cite{5d,asv}. We describe a product of two
curvatures as a state \be {\mathcal R}(Y) {\mathcal R}(Z) |0\rangle
\ee in a Fock space generated from the Fock vacuum $|0\rangle$
annihilated by the operators
\be
\bY^A_{i} |0\rangle = \bZ^A_{i} |0\rangle =0\,.
\ee
The annihilation operators $\bY^A_{i}$ and $\bZ^A_{i}$ have the
following commutation relations with $Y$ and $Z$
\be [\bY_{Ai} \,, Y^{Bj} ]= \delta_A^B \delta_i^j \q [\bZ_{Ai} \,,
Z^{Bj} ]=
\delta_A^B \delta_i^j\,,
\ee \ie $\bY_{Ai}$ and $\bZ_{Ai}$ are shorthand notations for
$\f{\partial}{\partial Y^{Ai}}$ and $\f{\partial}{\partial Z^{Ai}}$,
 respectively.

We will look for the action of the form
\be \label{act}
S=\frac{1}{2} \int
\langle 0| F(\bY,\bZ) {\mathcal R}(Y) {\mathcal R}(Z) |0\rangle\,,
\ee
where $F$ is a $(D-4)$-form constructed from the one-form AdS
background vielbein field $E^A={\mathcal D} V^A$ and the compensator
$V^A$
\be \label{F}
 F(\bY,\bZ) = \epsilon^{F_1\ldots F_{D-4} ABCDE } E_{F_1} \ldots E_{F_{D-4}}\,
 V_A\, \bY_{Bi}\, \bY_C^i\, \bZ_{Dj}\bZ_{E}^j\, :\Phi (u,w,v):\,.
\ee
In eq. (\ref{F}) we use the following notation
\be \label{u} u= V^C \bY_C^i V^D \bZ_{Di}\,,
\ee
\be \label{w} w=
\bY_{Di} \bZ^{Di}\,,
\ee
 \be \label{v}
v= \D_Y \D_Z\q \D_Y=
\f{1}{t(Y)} Y_C^k Y^{Cl} \bY_{Dk}
\bY_{l}^D\q \D_Z= \f{1}{t(Z)} Z_C^k Z^{Cl}
\bZ_{Dk} \bZ_{l}^D\,.
\ee
Note that the operators $\Delta_Y$, $\Delta_Z$ and, hence, $v$ are
well defined on the space $T$ of rectangular Young tableaux as one
can easily check using the decomposition (\ref{h}) and the property
(\ref{fc}). Indeed, the operator $\Delta(Y)$ decreases by one unit a
power of $t(Y)$ in the decomposition (\ref{h}). In particular, it
gives zero when acting on the traceless polynomials that correspond
to $p=0$ in (\ref{h}).

The normal ordering in (\ref{F}) is defined such that $v$ acts
before $u$ and $w$
\be
:\Phi (u,w,v): =\sum_{p=0}^\infty \Phi_p (u,w) v^p\,.
\ee
A normal ordering prescription is required because $v$ does not
commute with $u$ and $w$. In the sequel we omit the normal ordering
symbol.

The respective roles of the variables $u, w$ and $v$ are as follows.
The dependence of $\Phi$ on  $u$ (\ref{u}) takes care of the
projection of the higher--spin curvatures along a certain number of
$V^A$ similar to (\ref{VY}). The dependence of $\Phi$ on  $w$
(\ref{w}) controls the terms with different numbers of $O(2,D-1)$
covariant contractions between the higher--spin curvatures. The
dependence of $\Phi$ on  $v$ (\ref{w}) governs the coefficients for
different irreducible fields in the reducible system. Since $v$ acts
trivially on the traceless polynomials that correspond to the
irreducible higher--spin system, the dependence on $v$ is irrelevant
for their analysis, so in the irreducible case one can set $v=0$.

The condition (\ref{trc}) on the vectors in the triplet space
$T\otimes T $ is equivalent to \be \label{triv} \bY_A^i \bY^{Aj}
|\phi (Y,Z)\rangle  = Y_A^i Y^{Aj} \D_Y |\phi(Y,Z)\rangle\q \bZ_A^i
\bZ^{Aj} |\phi (Y,Z)\rangle  = Z_A^i Z^{Aj} \D_Z |\phi(Y,Z)\rangle\,.
\ee
An important property of the construction  is that if some $|h
(Y,Z)\rangle $ satisfies the triplet condition (\ref{triv}), $\D_Y
h(Y,Z)$, $\D_Z h(Y,Z) $ and, hence, $v h(Y,Z)$ also does. This is a
simple consequence of the fact that once $h(Y)$ has the form
(\ref{h}), then $\D_Y h(Y)$ also has this form.

The symmetry property of the action under the exchange of ${\mathcal
R}(Y)$ and ${\mathcal R}(Z)$ implies that
\be \label{sym}
\Phi(u,w,v) = \Phi (-u,-w,v)\,.
\ee

Being constructed from the gauge invariant curvatures, the action
(\ref{act}) is manifestly invariant under the higher--spin gauge
transformations. Consider a general variation of the action
\be
\delta S = \int \langle 0| F {\mathcal D} \,\delta \Omega (Y)\, R
(Z)\,|0\rangle\,.
 \ee
 Integrating by parts and taking into
account that $F$ in (\ref{F}) is constructed of manifestly
$o(2,D-1)$ covariant objects, we find that, in accordance with
(\ref{cv}),
 nonzero contributions to
the variation come only from the differentiation of the compensator
field $V^A$ that enters $F$ both directly and via $u$ (\ref{u}). The
resulting expression, obtained with the help of the identity
(\ref{idep1}), has the form
\be \label{var1}
\delta S = \frac{2}{D-3}\int  \langle 0| U \,\delta \Omega (Y)\, R
(Z)\,|0\rangle\,,
\ee
where \be \label{var2} U= \epsilon^{ABC} \bY_{Ai}
\bY^{i}_B \bZ_{Cj} \Big (V^E \bZ_{E}^j ((D-3) \Phi +2 u\f{\partial
\Phi}{\partial u} ) + (\bZ^{Gl} \bZ_G^j V^D \bY_{Dl} + \bY^G_l
\bZ_G^j V^D \bZ_{D}^l) \f{\partial \Phi}{\partial u}\Big )  +
\bY\leftrightarrow \bZ \ee
and
\be
 \epsilon^{ABC}\equiv \epsilon^{F_1\ldots F_{D-4} D ABC } E_{F_1} \ldots
 E_{F_{D-4}}V_D\,.
\ee

Our aim is to find such a function $\Phi(u,w,v)$ that the variation
of the action is identically zero for all the fields in the allowed
class except for the higher--spin vielbein and connection,
identified, respectively with the $V$-longitudinal and
$V$-transversal components of $(V_A\bY^A_2)^{s-2}\Omega(Y)|0\rangle$
(cf eq. (\ref{VY})). This condition, usually referred to as the
extra field decoupling condition, guarantees that the action is free
of higher derivatives carried by extra fields upon imposing
appropriate constraints that express them in terms of derivatives of
the higher--spin vielbein.

To this end it is helpful to use specific identities that hold as a
consequence of the properties of the class of fields under
consideration. The simplest of such identities follows from the
condition (\ref{sp2}) that the fields are $sp(2)$ singlets as they
describe rectangular Young tableaux. Namely, from the identity
\be\label{id1}
\langle 0| \epsilon^{ABC} \bY_{Ai} \bY^{i}_B
\bZ_{C}^j V^E \bZ_{E}^k [Y^F_{(j}\bY_{F k)} \,,\Lambda (u,w,v)]
\,\delta \Omega (Y)\, R (Z)\,|0\rangle =0
\ee
for any $\Lambda(u,w,v)$, where we also use that $\bY^A_i$ commutes
with $Z^A_i$, it follows that
\be \langle 0|
-\frac{2}{D-3}\epsilon^{ABC} \bY_{Ai} \bY^{i}_B \bZ_{C}^j \Big (
u\f{\partial
\Lambda}{\partial u} V^F \bZ_{Fj} + 2\bY_{F(j} \bZ^F_{l)}
\f{\partial \Lambda}{\partial w} V^G \bZ^l_G \Big )
\,\delta \Omega (Y)\, R (Z)\,|0\rangle =0\,.
\ee

A  more complicated identity, that follows from the conditions
(\ref{triv}), is
\bee \label{ntriv} &&-\frac{2}{D-3}\langle 0| \epsilon^{ABC} \bY_{Ai}
\bY^{i}_B \bZ_{C}{}_j \Big ( \bZ^{Ej} \bZ_E^k V^F \bY_{Fk}
(v\frac{\partial^4 W}{\partial w^4} -W)\nn\\
&+& V^F\bZ_F^j \Big ( 6 \frac{\partial^3}{\partial w^3} -3
\frac{\partial^3}{\partial u\partial w^2} + u(2
\frac{\partial}{\partial w} -\frac{\partial}{\partial u
})\frac{\partial^3}{\partial u\partial w^2} \Big)v W \Big )\,\delta
\Omega (Y)\, R (Z)\,|0\rangle =0\,
\eee
for any $W(u,w,v)$.

Summing  the variation (\ref{var1}) and (\ref{var2}) with the
identities (\ref{id1}) and (\ref{ntriv}) and considering
 the terms in front of
$$\epsilon^{ABC}
\bY_{Ai}\bY_{B}^i \bZ_{Cj} V^{F}\bZ_{F}^j\q \epsilon^{ABC}
\bY_{Ai}\bY_{B}^i \bZ_{Cj} \bZ_{F}^j \bZ^{{F}k} V^E \bY_{Ek}\q
\epsilon^{ABC} \bY_{Ai}\bY_{B}^i \bZ_{Cj} \bZ_E^{(j} \bY^{Ek)} V^E
Z_{{F}k}
$$
we obtain the three conditions
\be
\label{1} \ls\!\!\!\!\! (D-3) \Phi +2 u\f{\partial \Phi}{\partial u}
+\frac{1}{2} w \frac{\partial \Phi}{\partial u} +u\frac{\partial
{\Lambda}}{\partial u} -\Big ( 6 \frac{\partial^3}{\partial w^3} -3
\frac{\partial^3}{\partial u\partial w^2} + u(2
\frac{\partial}{\partial w} -\frac{\partial}{\partial u
})\frac{\partial^3}{\partial u\partial w^2} \Big)v {W} =2A(u,w,v),
\ee
\be \label{2} \f{\partial \Phi}{\partial u} +
{W} - v\f{\partial^4}{\partial w^4} {W}=0\,,
\ee
\be \label{3} \f{\partial \Phi}{\partial u} +2\f{\partial
{\Lambda}}{\partial w}=0\,,
\ee
where $A(u,w,v)$ determines the coefficients of the variation of the
action.

To obey the extra field decoupling condition the coefficient
function $A(u,w,v)$ should only depend on $uw$ and $v$
\be
\label{auw} A(u,w,v)=
\sum^\infty_{s=2}\sum_{p=0}^{[\frac{s}{2}]} A_{s,
p}(uw)^{s-2-4p}(v)^p\,,
\ee
where the dependence of $A_{s,p}$ on $s$ encodes relative
coefficients for the triplet actions with different spins while the
dependence on $p$ encodes the relative coefficients of the
irreducible spins within a given triplet system. Recall that the
property that the maximal number (\ie $s-2$) of indices of the
components $\delta \Omega$ or $R$ in (\ref{var1}) are contracted
with the compensators $V^A$ just implies that $A(u,w,v)$ depends on
$uw$, thus ensuring that the action depends only on the higher--spin
vielbein and connection.

The system of differential equations (\ref{1}), (\ref{2}) and
(\ref{3}) can be solved exactly. Indeed, (\ref{3}) implies that
\be \label{rel11} \Phi = 2\frac{\partial \varphi}{\partial w}\q
{\Lambda} = -\frac{\partial \varphi}{\partial u}\,. \ee Setting also
\be \label{rel21} {W} = \frac{\partial H
}{\partial u}\q \varphi =\psi +\frac{1}{2} \frac{\partial^3
H}{\partial w^3} v
\ee
we find by virtue of (\ref{2}) that \be
\label{rel31} H=-2\frac{\partial \psi}{\partial w} \ee and therefore
everything is expressed in terms of $\psi$ that has to satisfy the
equation (\ref{1}).

To analyze the resulting differential equation it is convenient to
introduce the following integral transform
\be \label{exp} \psi(u,w,v)= \oint d\s d\t e^{\s
u+\t w} \tilde{\psi} (\s,\t,v)\,, \quad A(u,w,v)= \oint d\s d\t
e^{\s u+\t w} \tilde{A} (\s,\t,v)\,,
\ee
where the integration measure is defined such that
\be \label{res}\oint d\s d\t
\s^p \t^q = \delta_{p+1}^{0}
\delta_{q+1}^{0}\q  p, q \in \mathbb{Z}\,.
\ee
Clearly, this transform relates the power series expansions as
follows
\be
\phi(u,w)=\sum_{n,m=0}^\infty a_{n,m} u^n w^m\quad\longleftrightarrow
\quad \tilde{\phi}(\s,\t)=\sum_{n,m=0}^\infty n!m! a_{n,m} \s^{-n-1}
\t^{-m-1}\,,
\ee
thus adding (removing) factorials to the coefficients. Let us stress
that, by analogy with the usual Cauchy integral, functions like
$\tilde{\psi} (\s,\t,v)$ and $\tilde{A} (\s,\t,v)$, that are
analytic in $\s$ and/or $\t$, do not contribute under the integral
(\ref{exp}). In the sequel, the equalities up to such terms will be
denoted by $\simeq$. The functions expandable in strictly negative
powers of $\s$ and $\t$ will be called relevant, while those
analytic in $\s$ and /or $\t$ will be called irrelevant. Thus
$\simeq$ is the equality modulo irrelevant functions.

By the transform (\ref{exp}), the equation (\ref{1}) amounts to
\bee \label{meq} && \ls\ls\Big (
(1-\t^4 v ) \Big((D-5)\t + \frac{1}{2} \s -2\s\t \frac{\partial
}{\partial \s} +
\frac{1}{2}
\s^2 \frac{\partial }{\partial \s}- \frac{1}{2} \s\t \frac{\partial
}{\partial \t} \Big )
\nn\\
&& + 4\t^4\s v + ( \s^3 \t^3 -2\t^4 \s^2)v \frac{\partial }{\partial
\s}\Big ) \tilde{\psi} (\s,\t,v)\simeq\tilde{A}(\s,\t,v)\,.
\eee

The following comment is now in order. The reason why we have chosen
the action (\ref{act}) with the function $F$ (\ref{F}) that depends
on the variables $u$ and $w$ is that this choice leads to the
first-order equation (\ref{meq}). This choice of variables differs
from that used by Alkalaev in \cite{alkal}, that has the advantage
of being  manifestly $sp(2)$ invariant,  allowing to avoid using the
identities (\ref{id1}) in the analysis. However, this is achieved at
the cost that the variables of \cite{alkal} are quartic in $\bY$ and
$\bZ$ (in our notation). This higher nonlinearity of the variables
of \cite{alkal} is expected to lead to nonlinearity of the identity
(\ref{ntriv}) to be translated to a higher-order counterpart of the
equation (\ref{meq}). Still, an interesting problem for the future
study is to reconsider the problem using the variables of
\cite{alkal}.

{}From eqs. (\ref{rel11})-(\ref{exp}) it follows that
\be
\tilde{\Phi}(\s,\t,v) \simeq 2\t (1-\t^4 v) \tilde{\psi}(\s,\t,v)\,.
\ee
The resulting equation on $\tilde{\Phi}$ is most conveniently
formulated in terms of the  variables
\be
\mu = \s\t\q \nu=2\t^2\,.
\ee
Using notation
\be
{\Phi}^\prime (\mu,\nu,v) =\tilde{\Phi} (\s,\t,v)\q {A}^\prime
(\mu,\nu,v) =\tilde{A} (\s,\t,v)\,,
\ee
the final equation is
\bee
\label{meq1} 
\Big (\frac{D-5}{2}+ L_0 +L_1\Big )
 {\Phi}^\prime (\mu,\nu,v)\simeq{A}^\prime(\mu,\nu,v)\,,
\eee
where
\be
L_0 = \frac{\mu}{\nu} -\mu (\frac{\partial}{\partial \mu}
+\frac{\partial}{\partial \nu})\,,
\ee
\be
\label{l1}
L_1=\frac{1}{2} v \Big ( 3\nu\mu \ + \mu^2 (\mu-\nu)
\frac{\partial}{\partial \mu}\Big )
\Big ( 1- \frac{1}{4} \nu^2 v\Big )^{-1}\,.
\ee

In principle, it is not hard to solve the equation (\ref{meq1}) with
the strict equality instead of $\simeq$ (see Appendix II).
 However, this method is
not most efficient  just because the right hand side of (\ref{meq1})
is known up to irrelevant terms. The formal solution obtained this
way for the function $\tilde{A} (\s,\t,v)$, that corresponds to
(\ref{auw}), leads to the wrong physical solution with an  unwanted
contribution at the boundary of the ``relevance region". That this
can happen follows for example from the $\frac{\mu}{\nu}$ term in
(\ref{meq1}), that can map irrelevant functions to the relevant
ones, thus giving a fake contribution at the boundary of the
``relevance region". To get rid of these unwanted terms, one has to
adjust an
 irrelevant right hand
side of (\ref{meq1}) such that it gives the relevant solution
$\Phi^\prime$ \footnote{In practice, it is enough to get rid of the
nonzero terms at the boundary of the set of irrelevant functions,
\ie those that are constants in $\s$ or $\t$.}. The solution to this
problem is not obvious, however. Hence, we proceed differently, by
solving the equation (\ref{meq1}) via an appropriate
 Ansatz directly in the relevant class.

Note that in terms of the variables $\mu$ and $\nu$, the relevant
functions have the form
\be
\label{class}
F^{rel}(\mu,\nu) = \mu^{-1} P(\mu^{-1},\nu^{-1})
\ee
for an entire function $P(x,y)$ (polynomial for a given triplet
system).

The key observation is that the following identity is true for $a>
-1$ and any  $C(x,y)$
\bee
\label{ke}
&&(L_0 +a)\Big(\mu^{-1}\int_0^1 ds \int_0^1 dt (1-t)^{a}
C(s(1-t)\mu^{-1} -(1-s) t\nu^{-1}, (1-s)\nu^{-1} )\Big )\nn\\&&
=\mu^{-1}\int_0^1 ds
 C(s \mu^{-1}, (1-s)\nu^{-1} ) -
\nu^{-1} \int_0^1 dt C(- t\nu^{-1}, \nu^{-1} )\,,
\eee
which follows from the following two elementary identities
\be
L_0 \mu^{-1} = \mu^{-1} (L_0+1)
\ee
and
\bee
&&(L_0 +a) \Big(\int_0^1 ds \int_0^1 dt (1-t)^{(a-1)}
C(s(1-t)\mu^{-1} -(1-s) t\nu^{-1}, (1-s)\nu^{-1} )\Big )=
\nn\\
&&-\int_0^1 ds \int_0^1 dt \Big ( \frac{\partial}{\partial t}\big(
(1-t)^a C(s(1-t)\mu^{-1} -(1-s) t\nu^{-1}, (1-s)\nu^{-1} )\big)
\nn\\&& + \frac{\mu}{\nu}
\frac{\partial}{\partial s}\big((1-s)C(s(1-t)\mu^{-1} -(1-s) t\nu^{-1},
(1-s)\nu^{-1} )\big)\Big )\,.
\eee

Now we observe that the second term on the right hand side of
(\ref{ke}) is irrelevant because it is $\mu$ independent and, hence,
$\s$--independent. (Note that this is just the irrelevant term to be
added to make it possible to reconstruct an appropriate formal
solution of (\ref{meq1}) as discussed above.)
 Hence, the identity (\ref{ke}) gives
\be
\label{key}
(L_0 +a) \Big(\mu^{-1}\int_0^1 ds \int_0^1 dt
(1-t)^{a} C(s(1-t)\mu^{-1} -(1-s) t\nu^{-1}, (1-s)\nu^{-1} )\Big )=
U(C)(\mu,\nu)
\,,
\ee
where
\be
\label{ac}
 U(C)(\mu, \nu) \equiv \mu^{-1}\int_0^1 ds
 C(s \mu^{-1}, (1-s)\nu^{-1} )\,.
\ee
Using that
\be
\label{beta}
\int_0^1 dt t^a (1-t)^b = \frac{a!b!}{(a+b+1)!}\,,
\ee
 we obtain
\be
C(\mu^{-1},\nu^{-1}) = \sum_{n,m=0}^\infty c_{n,m}\,
\mu^{-n}\nu^{-m}
\quad\longrightarrow\quad
U(C)(\mu,\nu) = \sum_{n,m=0}^\infty
\frac{n!\,m!}{(n+m+1)!}c_{n,m}\, \mu^{-n-1}\nu^{-m}\,.
\ee
The inverse transform to (\ref{ac}) can be written in the form
\be
C(\mu^{-1},\nu^{-1}) = U^{-1}(A)(\mu^{-1},\nu^{-1})\equiv -\mu^2
\frac{\partial}{\partial \mu}\oint d\s^\prime d\t^\prime
\frac{1}{\mu^\prime}
A( (\mu^{-1} +\nu^{-1}\nu^\prime\mu^\prime{}^{-1})^{-1},
 (\mu^{-1}\mu^\prime\nu^\prime{}^{-1} +\nu^{-1})^{-1})\,.
\ee

Now we can write the solution of the equation
\be
\label{eq0}
(L_0 +a ) \tilde{\Phi}_0 (\mu,\nu) =A_0^\prime (\mu,\nu)
\ee
in the form
\be
\label{eq01}
\tilde{\Phi}_0 (\mu,\nu) =(L_0 +a)^{-1} A_0^\prime (\mu,\nu)\,,
\ee
where
\be
\label{int1}
(L_0 +a)^{-1} A_0^\prime (\mu,\nu) =
\mu^{-1}\int_0^1 ds \int_0^1 dt (1-t)^{a}
U^{-1}(A)(s(1-t)\mu^{-1} -(1-s) t\nu^{-1}, (1-s)\nu^{-1})\,.
\ee

An important property of all maps under consideration is that they
act within the class of relevant functions. Also
 it is clear that if $A_0^\prime (\mu,\nu)$
is a homogeneous function of degree $n$, then the same is true for
$\tilde{\Phi}_0 (\mu,\nu)$. This property manifests  that the
solutions for different spins, corresponding to different
homogeneity  degrees, are independent.

To obtain the formula for the solution of (\ref{meq1}), that
determines the coefficients of the action, in terms of the expansion
in powers of $v$, which is equivalent to the lower-spin expansion
within the system of triplet fields, it remains to define precisely
the multiplication law by $\mu$ and $\nu$ on the functions of the
class (\ref{class}). The rule is simply that the irrelevant terms
should be discarded. This means that $\mu P(\mu^{-1}, \nu^{-1})$
should be replaced by $\mu\circ P(\mu^{-1}, \nu^{-1})=\mu
(P(\mu^{-1}, \nu^{-1})- P(0, \nu^{-1}))$. Equivalently, one can
write
\be
\mu\circ P(\mu^{-1}, \nu^{-1}) = \int_0^1 dt  P_{1,0}(t\mu^{-1}, \nu^{-1})\q
\nu\circ P(\mu^{-1}, \nu^{-1}) = \int_0^1 dt P_{0,1}(\mu^{-1}, t\nu^{-1})\,,
\ee
where
\be
P_{n,m}(x,y) = \frac{\partial^{n+m}}{\partial x^n\partial
y^m}P(x,y)\,.
\ee

Successive application of this formula gives
\be
\label{circ1}
(\mu^n\nu^m)\circ P(\mu^{-1}, \nu^{-1}) = \frac{1}{nm}
\int_0^1 dt (1-t)^{n-1}\int_0^1 du (1-u)^{m-1}
 P_{n,m}(t\mu^{-1}, u\nu^{-1})\q m,n>0\,.
\ee

The solution of the equation (\ref{meq1}) can now be written in the
form
\be
\label{feq}
 {\Phi}^\prime (\mu,\nu,v) =\sum_{n=0}^\infty (-1)^n
\Big(L_0+\frac{D-5}{2}\Big)^{-1}\Big( L_1\Big(L_0+\frac{D-5}{2}\Big)^{-1}\Big)^n\,
{A}^\prime(\mu,\nu,v)\,,
\ee
where all multiplications with $\mu$ and $\nu$ contained in
 $L_1$ (\ref{l1}) should be understood as the $\circ$--multiplication
 (\ref{circ1}). (Note that to work from the very beginning within the relevant class,
 one should replace usual multiplication by $\circ$ directly in (\ref{meq1}),
 which however then becomes an integro-differential equation.)

To find the action for the triplet system we have to apply the
formula (\ref{feq}) to ${A}^\prime (\mu,v)$ because the functions
$\tilde{A}(\s,\t,v)=\s^p\t^q \rho(v)$ with $p< q$ give rise to a
trivial variation by virtue of the Young symmetry properties of the
fields, while those with $p> q$ give rise to the actions that
contain extra fields, thus leading to the field equations with
higher derivatives. In accordance with (\ref{auw}), for the spin $s$
triplet system
\be \label{dynan} \tilde{A}
(\mu,v)=\mu^{1-s} \tilde{A}_s (v)\,,
\ee
where the function $\tilde{A}_s (v)$ determines the coefficients of
the action for the irreducible spin components in the triplet system
with the highest spin $s$. The most convenient choice is
$\tilde{A}_s (v)=\tilde{A}_s$, \emph{ i.e.}
\be
\label{As} \tilde{A} (\mu,v)=\mu^{1-s}{A}_s\,,
\ee
where $A_s$ is an overall normalization constant coefficient for the
reducible spin--$s$ system.

The absence of the extra fields in the triplet action (\ref{act}),
the Young--tableau structure and symmetry properties of its
components allow us to conclude that in the gauge
$V_A=\frac{1}{\sqrt{-\Lambda}}\,\delta^{0'}_A$ it reduces to the
bosonic triplet action (\ref{AdSactiongf}).

\subsection{Irreducible case}

To illustrate the obtained result let us consider the example of an
irreducible massless field. In this case, we should set $v=0$ since
$v$ acts trivially on the irreducible $\Omega$.

The covariant action for the irreducible case was obtained in
\cite{5d} in the form
\bee
\label{act5d}
S=&&\frac{1}{2}\sum_{p=0}^{s-2}  \alpha(s) 2^p
\frac{(p+1)(\frac{d-5}{2} +p)!}{p!}
 \epsilon^{F_1\ldots F_{D-4} ABCDE } E_{F_1} \ldots E_{F_{D-4}}\,
 V_A\,V^{H_1}\ldots V^{H_2p}\nn\\
&& R_{BH_1\ldots H_p G_1\ldots G_{s-p-2}\, CF_1\ldots F_{s-2}}\wedge
R_{DH_1\ldots H_{p}}{}^{G_1\ldots G_{s-p-2}}{}_{,\,E}{}^{F_1\ldots
F_{s-2}}\,
\eee
with some  overall spin-dependent normalization factor $\alpha (s)$.
The coefficients in this action were determined in \cite{5d} from
the extra field decoupling condition that guarantees that it
properly describes a spin $s$ irreducible massless field.

The components of the higher--spin curvatures in (\ref{act5d})
result from the following expansion of the higher--spin curvatures
(\ref{RY})
\be
\label{rex}
R(Y)=\sum_{s=1}^\infty \frac{1}{((s-1)!)^2} R^{A_1\ldots
A_{s-1},B_1\ldots B_{s-1}}
 Y^1_{A_1}\ldots Y^1_{A_{s-1}} Y^2_{B_1}\ldots Y^2_{B_{s-1}}\,.
\ee
To compare the action (\ref{act}), (\ref{F}) with (\ref{act5d}) let
us evaluate $\langle 0|u^p w^{2s^\prime -p} R(Y)R(Z)|0\rangle$.
Using the expansion (\ref{rex}) along with $u$ (\ref{u}) and $w$
(\ref{w}) in the form
\be
u=-V_A V_B \left (\frac{\partial^2}{\pa Y_A^1 \pa Z_B^2} -
\frac{\partial^2}{\pa Y_A^2 \pa Z_B^1}\right )\q
w= \left (\frac{\partial^2}{\pa Y_A^1 \pa Z^{A2}} -
\frac{\partial^2}{\pa Y_A^2 \pa Z^{A1}}\right )\,,
\ee
direct differentiation gives
\bee
{}&&\langle 0|u^p w^{2s^\prime -p} R(Y)R(Z)|0\rangle
=\sum_{n,m=0}^\infty (-1)^{p+s^\prime}\delta (n+m-s^\prime)
\frac{p!(2s^\prime -p)!}{n!(p-n)!m!(2s^\prime -p -m)!}V_{C_{1}}\ldots V_{2p-n}\nn\\
&& R^{C_1\ldots C_n A_1\ldots A_m,}{}^{C_{n+1}\ldots C_{p-n}
B_1\ldots B_{s^\prime -p +n}}\, R^{C_{p-n+1}\ldots
C_{2(p-n)}}{}_{B_1\ldots B_{s^\prime -p +n}, A_1\ldots
A_m}{}^{C_{2(p-n)+1}\ldots C_{2p-n}}\,.
\eee
Then using repeatedly the Young properties of the higher--spin
curvatures in the form
\bee
&&R_{A_1\ldots A_{s^\prime -n} (B_1\ldots B_n,B_{n+1}\ldots B_{n+k})
C_1\ldots C_{s^\prime -n}}\nn\\&&=(-1)^k\frac{n!(s^\prime
-n)!}{(n-k)!(s^\prime -n-k)!} R_{ B_1\ldots  B_{n+k}(A_1\ldots
A_{s^\prime -n-k}, A_{s^\prime -n-k+1}\ldots A_{s^\prime
-n})C_1\ldots C_{s^\prime -n}}
\eee
it is not difficult to obtain
\bee
{}&&\langle 0|u^p w^{2s^\prime -p} R(Y)R(Z)|0\rangle = (-1)^{p}(p+1)
\frac{(2s^\prime -p)!}{s^\prime!(s^\prime-p)!}
V_{C_{1}}\ldots V_{C_{2p}}\nn\\
&& R^{C_1\ldots C_p A_1\ldots A_{s^\prime -p} }{}_,{}^{ B_1\ldots
B_{s^\prime }}\, R^{C_{p+1}\ldots C_{2p}}{}_{A_1\ldots A_{s^\prime
-p},} {}_{B_1\ldots B_{s^\prime}}\,.
\eee
To apply this formula to the action (\ref{act}) it remains to
observe that the contraction of the indices $A,B,C,D$ in  the action
(\ref{act5d}) with the epsilon symbol does not
 change the Young symmetry properties with respect to the other
 indices carried by the higher--spin curvatures, shifting effectively
 the parameter $s^\prime$ by one unit.

Direct comparison shows that the action (\ref{act5d}) is reproduced
by the function $\Phi(u,w,0)$ in (\ref{F}) of the form
\be
\Phi(u,w,0)= \sum_{p,s} \alpha(s)(-2)^p \frac{(\frac{D-5}{2}+p)!
(s-p-2)!(s-2)!}{p! (2(s-2)-p)!} u^p w^{2(s-2) -p}\,.
\ee
The transform (\ref{exp}) gives
\be
\label{phcom}
\tilde{\Phi}(\nu,\mu,0) =\sum_{s=2}^\infty\sum_{p=0}^{s-2} \alpha(s)(-1)^p {\left(
\frac{D-5}{2}+p\right ) !
(s-p-2)!(s-2)!} \mu^{-p-1} \nu^{-(s-p-2)}\,.
\ee

On the other hand, choosing  $\tilde{A}(\mu,0)$, in accordance with
(\ref{As}), we obtain
\be
C(\mu^{-1},\nu^{-1}) = U^{-1}(A^\prime)(\mu^{-1},\nu^{-1})= (s-1)A_s
\mu^{2-s}
\ee
and, by (\ref{int1}) with $a=\frac{D-5}{2}$,
\be
\label{int00}
\tilde{\Phi}_0 (\mu,\nu)= (s-1)A_s
\mu^{-1}\int_0^1 du \int_0^1 dt (1-t)^{\frac{D-5}{2}}
(u(1-t)\mu^{-1} -(1-u) t\nu^{-1})^{s-2}\,.
\ee
Using (\ref{beta}) this gives
\be
\tilde{\Phi}(\nu,\mu,0) =
\sum_{s=2}^\infty\sum_{p=0}^{s-2}(-1)^{s+p} A_s(-1)^p \frac{
(\frac{D-5}{2}+p )! (s-p-2)!}{(\frac{D-5}{2}+s-2 )!} \mu^{-p-1}
\nu^{-(s-p-2)}\,.
\ee
We observe that this formula indeed coincides with (\ref{phcom})
provided that
\be
A_s =(-1)^s (\frac{D-5}{2}+s-2 )!(s-2)! \alpha(s)\,.
\ee

Expanding the expression (\ref{feq}) in powers of $v$ one can
systematically reconstruct the covariant action coefficients for the
lower--spin components of the triplet systems.

\section{Towards the AdS covariant formulation of the fermionic higher--spin fields}
\label{fermads}
\setcounter{equation}0
In conclusion of the main part of this paper we present basic
ingredients of the AdS covariant description of the relaxed systems
of fermionic massless symmetric higher--spin fields.

In the $o(2,D-1)$ covariant formalism, a single symmetric fermionic
massless field of spin $s$ is  described by a tensor--spinor
one--form $\Psi_{A_1\ldots A_{s -\frac{3}{2}}\,,B_1\ldots B_{s
-\frac{3}{2}}}$ that has properties of a gamma--transversal two--row
rectangular Young tableau
\be
\Psi_{(A_1\ldots A_{s -\frac{3}{2}}\,,A_{\s} )B_2\ldots B_{s -\frac{3}{2}}} =0\q
\Gamma^{A_1} \Psi_{A_1\ldots A_{s -\frac{3}{2}}\,,B_1\ldots B_{s -\frac{3}{2}}}=0\,.
\ee
 This is because the decomposition of this $spin(2,D-1)$
irreducible tensor-spinor into the $spin(1,D-1)$ Lorentz irreducible
tensor--spinors gives all gamma--transversal two--row Young
tableaux, which is precisely the pattern of higher-spin connections
introduced in
\cite{V87,Vasiliev:1987tk}.

Like in the bosonic case, an \emph{unrestricted} rectangular
two--row tensor-spinor can be described by  $Y$-dependent spinor
\be
\label{polyf}
f^{\hat\ga} (Y) = \sum_{n=0}^\infty f_{A_1\ldots A_n\,,B_1 \ldots
B_n}^{\hat\ga} Y^{A_1}_1 \ldots Y^{A_n}_1 Y^{B_1}_2 \ldots
Y^{B_n}_2\,,
\ee
that satisfies the $sp(2)$ invariance condition (\ref{sp2}) (here
$\hat\ga\,,\hat\gb$ are spinor indices of $o(2,D-1)$). Irreducible
$o(2,D-1)$--modules are described by polynomials
 $f_{A_1\ldots A_n\,,B_1 \ldots B_n}^{\hat\ga}$, that, in addition to
(\ref{sp2}), satisfy the gamma-transversality  condition
\be
\label{gtr}
\Gamma^A{}^{\hat\ga}{}_{\hat\gb}\frac{\partial}{\partial Y^A_i} f^{\hat\gb} (Y) =0\,.
\ee

The class of functions $T$ with the relaxed gamma--traceless
conditions appropriate for the description of fermionic higher--spin
triplets is formed by various polynomials of the form
\be
\label{f}
f(Y) = \sum_{p=0}^\infty (\Gamma (Y))^p f_p (Y)\,,
\ee
where $f_p (Y)$ satisfy the conditions (\ref{sp2}) and (\ref{gtr})
and
\be
\label{G}
\Gamma(Y) \equiv \Gamma_i \Gamma^i \,,\qquad \Gamma_i \equiv
\Gamma^A Y_{Ai}.
\ee

Note that $(\Gamma(Y))^p$ in (\ref{f}) is understood as the $p^{th}$
matrix power of $\Gamma^{\hat\ga}{}_{\hat\gb}(Y)$ (\ref{G}). It is
easy to see that the relation
$\Gamma^A\Gamma^B+\Gamma^B\Gamma^A=2\eta^{AB}$  implies
\be\label{eta}
\Gamma_i\Gamma_j+\Gamma_j\Gamma_i=2\eta_{ij}(Y)\q
\eta_{ij}(Y) = \eta_{AB}Y^A_i Y^B_j\,.
\ee
{}From here it follows that
\be\label{g2}
(\Gamma(Y))^2{}^{\hat\gb}{}_{\hat\ga} =
-2\,t(Y)\,\delta_{\hat\ga}^{\hat\gb}\,,
\ee
where $t(Y)=\eta_{ij} (Y)\eta^{ij}(Y)$ was introduced in (\ref{t}).
Note also that
\be\label{eta2}
\eta_{ki} \eta^{kj}=-\eta^k_{i} \eta^{j}_k=\frac{1}{2}\,\delta_i^j\,t(Y)\,.
\ee
Recall that the $sp(2)$ indices are raised and lowered by the
$sp(2)$ symplectic forms according to (\ref{symind}).

Since $i$ takes two values, $\Gamma(Y)$ plays the role of the
``$\Gamma_3$--matrix", \emph{i.e.} it anticommutes with $\Gamma_i$
\be\label{gamma3}
\Gamma(Y)\,\Gamma_i+\Gamma_i\,\Gamma(Y)=0\,,
\ee
which together with (\ref{eta}) implies a useful relation
\be\label{gg}
\Gamma(Y)\,\Gamma^i=2\eta^{ij}(Y)\,\Gamma_j\,.
\ee

We observe that $\Gamma(Y)$ satisfies the condition (\ref{sp2})
because the indices in (\ref{G}) are contracted in the $sp(2)$
invariant way.
 Therefore, $f(Y)$ of the form (\ref{f}) satisfies the two-row
Young symmetry condition (\ref{sp2}).

It is not hard to see that the characteristic property of the space
$T$ (\ref{f}) is that, taking into account (\ref{gg}),
 any its element satisfies the property
\be\label{fg}
f(Y)\in T\,:\qquad \Gamma^A\frac{\partial}{\partial Y^A_i} \,f(Y) =
\Gamma^i(Y)\, g(Y)\,,
\ee
where $g(Y)$ is again a $Y$--polynomial of the type (\ref{f})
(\emph{i.e.} $g(Y)\in T$).

{}From eqs. (\ref{fg}) and (\ref{gg}) it follows that the relaxed
gamma--transversality condition in $T$ has the form
\be
\label{ftrc}
\Gamma\Gamma^A\frac{\partial}{\partial Y^A_i} f(Y)  =
- 2\eta^i{}_j (Y)
\Gamma^A \frac{\partial}{\partial Y^A_{j}} f(Y)\,.
\ee
Equivalently, it can be written in the form
\be
\label{chirality}
P^i{}_j\,\Gamma^A\, \frac{\partial}{\partial Y^A_j}f(Y)=0\,,
\ee
where
\be
P^i{}_j = \delta^i_j -\frac{1}{2t(Y)} \Gamma \Gamma^i
\Gamma_j=\frac{1}{2}\left(
\delta^i_j -\frac{1}{t(Y)} \Gamma (Y)\eta^i{}_j(Y)\right )
\ee
is the projector, \emph{i.e.} $ P^i{}_jP^j{}_k =P^i{}_k\,. $

Clearly, the constraint (\ref{chirality}) is weaker than the
gamma--transversality condition (\ref{gtr}) satisfied by the
irreducible higher--spin fields. As expected, eq. (\ref{chirality})
singles out the reducible (triplet) systems of symmetric fermionic
fields: all  $o(2,D-1)$ irreducible components of a  degree
$2(s-3/2)$ one--form connection $\Psi(Y)$
 describe two-row rectangular gamma-transversal
tensor--spinors that correspond to the set of fermionic fields of
spins $s, s-1, s-2,\cdots,
\frac{3}{2}$. Thus, the $T$--valued one--form spinors
$\Psi(Y)$ which are subject to the gauge transformations
\be
\delta \Psi (Y)={\mathcal D}\, \xi(Y), \qquad {\mathcal D}\,{\mathcal D}=0
\ee
describe in the AdS $o(2,D-1)$--covariant way the fermionic
higher--spin triplets discussed in detail in Subsection \ref{ftads}.

The manifestly gauge and $o(2,D-1)$--invariant analysis of the
fermionic action is more involved than of the bosonic one. Even the
case of irreducible higher--spin fermions is currently under
investigation \cite{sh}. We therefore leave for the future
consideration the formulation of the manifestly gauge and
$o(2,D-1)$--invariant action for the fermionic triplet system.

\section{Conclusion}
We have considered the frame--like Lagrangian formulation of free
systems of bosonic and fermionic higher--spin fields in flat and AdS
backgrounds of arbitrary dimension. We have shown that the
higher--spin systems described by an unconstrained higher--spin
vielbein and by the connections which are subject to weaker
(gamma)--trace constraints than those required for the description
of single Fronsdal and Fang--Fronsdal fields correspond to the
higher--spin triplets whose fields are associated with certain
components of the higher--spin vielbein and connection. We have thus
endowed the triplet fields with a clear geometrical meaning. This
allowed us to identify the appropriate form of the gauge
transformations of the fermionic triplets in AdS space and construct
the gauge invariant action which describes their dynamics. We have
also shown how upon imposing the pure gauge constraints on the
(gamma)--trace of the higher--spin vielbeins one reduces the triplet
systems to the frame--like versions of the unconstrained
formulations of single higher--spin fields considered in
\cite{Buchbinder:2007ak,Francia:2007ee}.

An interesting direction of future research is the extension of the
obtained results to the interacting level. An important related
question is whether the conditions (\ref{tr}) and (\ref{ftrc}) have
an algebraic meaning that would allow one to figure out what might
be a higher--spin algebra underlying these reducible higher--spin
multiplets.

\subsection*{Acknowledgements}
We are grateful to Konstantin Alkalaev, Augusto Sagnotti and Mirian
Tsulaia for fruitful discussions. We would also like to thank Dieter
Luest for kind hospitality extended to us at
Ludwig--Maximilians--Universitaet, Muenchen where part of this work
was done. This work was partially supported by the INTAS Project
Grant 05-1000008-7928. Work of D.S. was partially supported by the
European Commission FP6 program MRTN-CT-2004-005104 ``Forces
Universe" and by the MIUR Research Project PRIN-2005023102 in which
D.S. is associated to the Department of Physics of Padova
University, by the INFN Special Initiatives TS11 and TV12, and by
research grants from the Spanish Ministry of Education and Science
(FIS2005-02761). Work of M.V. was partially supported by RFBR Grant
No 05-02-17654, LSS No 4401.2006.2 and the Alexander von Humboldt
Foundation Grant PHYS0167.

\def\thesection{}
\def\theequation{A.\arabic{equation}}
\section{Appendix A}
\setcounter{equation}0

In this Appendix we show that the variation of the action
(\ref{fract}) with respect to the connection $\omega$ results in the
zero--torsion condition (\ref{zt}) both in the Fronsdal case and
(modulo spin 1) in the case in which the higher--spin vielbein is
traceful and the connection $\omega$ is subject to the relaxed trace
condition (\ref{trnn}).

\textbf{In the \FR case}, the \emph{tracelless} higher--spin vielbein $\tilde e_{m;\, n_1
\cdots n_{s-1}}$ contains the irreducible
Lorentz tensors described by the following Young tableaux
\be
\begin{picture}(6,7)(0,0)
\multiframe(0,0)(7.5,0){1}(7,7){}
\end{picture}\
\otimes
\begin{picture}(50,12)(0,0)
\multiframe(0,0)(13.5,0){1}(30,7){}\put(33,1){\tiny{$ s-1$}}
\end{picture}\ =
\begin{picture}(40,12)(0,0)
\multiframe(0,0)(13.5,0){1}(30,7){}\put(33,1){\tiny{$ s$}}
\end{picture}\oplus
\begin{picture}(50,12)(0,0)
\multiframe(0,0)(13.5,0){1}(30,7){}\put(33,1){\tiny{$ s-2$}}
\end{picture}\oplus
\begin{picture}(55,15)(-5,5)
\multiframe(0,0)(13.5,0){1}(7,7){}\put(9,1){{\tiny $ 1$}}
\multiframe(0,7.5)(13.5,0){1}(35,7){}\put(38,9){\tiny{$ s-1$}}
\end{picture}\,\,.
\label{e}\ee
The first tableau of length $s$ on the right hand side of (\ref{e})
describes the totally symmetric and traceless part of $\tilde e$,
the second tableau of the length $s-2$ corresponds to the traceless
$\eta^{mn_{s-1}}\tilde e_{m;\, n_1 \cdots n_{s-1}}$ and the hook
tableau corresponds to the irreducible (traceless) part of $\tilde
e$ that satisfies $\tilde e_{(m;\, n_1\cdots n_{s-1})}=0$.

Because of the gauge transformation law of the higher--spin vielbein
\be\label{deviel1}
\delta\,\tilde e^{a_1\cdots a_{s-1}}
=d\tilde \xi^{a_1\cdots a_{s-1}}-d x^m\,\tilde\xi^{a_1\cdots
a_{s-1},b}\eta_{mb}\,,
\ee
the hook part of the vielbein can be gauged to zero by the gauge
shift with the parameter $\tilde\xi^{a_1\cdots a_{s-1},b}$. As a
result, the remaining part of the vielbein is the combination of two
totally symmetric traceless tensor of rank $s$ and $s-2$ equivalent
to the double traceless Fronsdal field.

The fact that the zero torsion condition (\ref{ofc11}) is equivalent
to the equation of motion (\ref{eeq1}) is deduced as follows.

The torsion tensor $T_{mn;n_1 \cdots n_{s-1}}$ contains the
irreducible (traceless) components describe by the following set of
Young tableaux
\be
\begin{picture}(09,13)
{
\put(05,05){\line(1,0){05}}%
\put(05,10){\line(1,0){05}}%
\put(05,0){\line(1,0){05}}%
\put(05,0){\line(0,1){10}}%
\put(10,0.0){\line(0,1){10}}
}
\end{picture}\,\,
\otimes
\begin{picture}(50,12)(0,0)
\multiframe(0,0)(13.5,0){1}(30,7){}\put(33,1){\tiny{$ s-1$}}
\end{picture}\ =
\begin{picture}(55,15)(-5,10)
\multiframe(0,0)(13.5,0){1}(7,7){}\put(9,1){{\tiny $ 1$}}
\multiframe(0,7.5)(13.5,0){1}(7,7){}\put(9,8){\tiny{$ 1$}}
\multiframe(0,15)(13.5,0){1}(35,7){}\put(38,15){\tiny{$ s-1$}}
\end{picture}\,\,
\oplus
\begin{picture}(55,15)(-5,5)
\multiframe(0,0)(13.5,0){1}(7,7){}\put(9,1){{\tiny $ 1$}}
\multiframe(0,7.5)(13.5,0){1}(35,7){}\put(38,9){\tiny{$ s$}}
\end{picture}\,\,
\oplus
\begin{picture}(55,15)(-5,5)
\multiframe(0,0)(13.5,0){1}(7,7){}\put(9,1){{\tiny $ 1$}}
\multiframe(0,7.5)(13.5,0){1}(35,7){}\put(38,9){\tiny{$ s-2$}}
\end{picture}\,\,
\oplus
\begin{picture}(50,12)(0,0)
\multiframe(0,0)(13.5,0){1}(30,7){}\put(33,1){\tiny{$ s-1$}}
\end{picture}.
\label{TY}\ee

Note that the last two tableaux describe the two irreducible parts
of $\eta^{nn_1}\,T_{mn;n_1 \cdots n_{s-1}}$.

On the other hand the connection $\tilde\o_{l;}{}^{n_1 \cdots
n_{s-1},\,m}$ has the following decomposition
\be
\begin{picture}(6,7)(0,0)
\multiframe(0,0)(7.5,0){1}(7,7){}
\end{picture}\
\otimes
\begin{picture}(43,15)(-5,5)
\multiframe(0,0)(13.5,0){1}(7,7){}\put(9,1){{\tiny $ 1$}}
\multiframe(0,7.5)(13.5,0){1}(35,7){}\put(38,9){\tiny{$ s-1$}}
\end{picture}\,\,\,\,\,\,\,\, =
\begin{picture}(55,15)(-5,5)
\multiframe(0,0)(13.5,0){1}(14,7){}\put(16,1){{\tiny $ 2$}}
\multiframe(0,7.5)(13.5,0){1}(35,7){}\put(38,9){\tiny{$ s-1$}}
\end{picture}\,\,
\oplus
\begin{picture}(55,15)(-5,10)
\multiframe(0,0)(13.5,0){1}(7,7){}\put(9,1){{\tiny $ 1$}}
\multiframe(0,7.5)(13.5,0){1}(7,7){}\put(9,8){\tiny{$ 1$}}
\multiframe(0,15)(13.5,0){1}(35,7){}\put(38,15){\tiny{$ s-1$}}
\end{picture}\,\,
\oplus
\begin{picture}(43,15)(-5,5)
\multiframe(0,0)(13.5,0){1}(7,7){}\put(9,1){{\tiny $ 1$}}
\multiframe(0,7.5)(13.5,0){1}(35,7){}\put(38,9){\tiny{$ s$}}
\end{picture}\,\,
\oplus
\begin{picture}(55,15)(-5,5)
\multiframe(0,0)(13.5,0){1}(7,7){}\put(9,1){{\tiny $ 1$}}
\multiframe(0,7.5)(13.5,0){1}(35,7){}\put(38,9){\tiny{$ s-2$}}
\end{picture}\,\,
\oplus
\begin{picture}(50,12)(0,0)
\multiframe(0,0)(13.5,0){1}(30,7){}\put(33,1){\tiny{$ s-1$}}
\end{picture}.
\label{oY}\ee
Observe that the two decompositions (\ref{TY}) and  (\ref{oY})
differ by the first tableau on the right hand side of (\ref{oY})
which, however, is just the pure gauge part of the higher--spin
connection which can be set to zero by the gauge shift
\be\label{do2}
\delta\tilde\omega_{m;a_1\cdots a_{s-1},b_1}=-\tilde\xi_{a_1\cdots
a_{s-1},bm}\,.
\ee
As a result the torsion tensor has as many components as the
higher--spin connection $\tilde\o^{n_1
\cdots n_{s-1},\,m}$ modulo its pure gauge part. So the number of the independent field
equations of $\tilde\omega$ in (\ref{eeq1}) equals to the number of
the components of the torsion.

\textbf{In the triplet case} in which the vielbein is unconstrained while the connection
and gauge parameters satisfy the relaxed traceless conditions
(\ref{trnn}) and (\ref{trmn}), the torsion tensor has the
decomposition in terms of the
\emph{traceful} Young tableaux. It thus does not contain the last two
terms in (\ref{TY}), namely,
\be
\begin{picture}(09,13)
{
\put(05,05){\line(1,0){05}}%
\put(05,10){\line(1,0){05}}%
\put(05,0){\line(1,0){05}}%
\put(05,0){\line(0,1){10}}%
\put(10,0.0){\line(0,1){10}}
}
\end{picture}\,\,
\otimes
\begin{picture}(50,12)(0,0)
\multiframe(0,0)(13.5,0){1}(30,7){}\put(33,1){\tiny{$ s-1$}}
\end{picture}\ =
\begin{picture}(55,15)(-5,10)
\multiframe(0,0)(13.5,0){1}(7,7){}\put(9,1){{\tiny $ 1$}}
\multiframe(0,7.5)(13.5,0){1}(7,7){}\put(9,8){\tiny{$ 1$}}
\multiframe(0,15)(13.5,0){1}(35,7){}\put(38,15){\tiny{$ s-1$}}
\end{picture}\,\,
\oplus
\begin{picture}(55,15)(-5,5)
\multiframe(0,0)(13.5,0){1}(7,7){}\put(9,1){{\tiny $ 1$}}
\multiframe(0,7.5)(13.5,0){1}(35,7){}\put(38,9){\tiny{$ s$}}
\end{picture}.
\label{TY3}\ee

 The \emph{traceful} Young tableau decomposition of the
connection is
\be
\begin{picture}(6,7)(0,0)
\multiframe(0,0)(7.5,0){1}(7,7){}
\end{picture}\
\otimes
\begin{picture}(43,15)(-5,5)
\multiframe(0,0)(13.5,0){1}(7,7){}\put(9,1){{\tiny $ 1$}}
\multiframe(0,7.5)(13.5,0){1}(35,7){}\put(38,9){\tiny{$ s-1$}}
\end{picture}\,\,\,\,\,\,\,\, =
\begin{picture}(55,15)(-5,5)
\multiframe(0,0)(13.5,0){1}(14,7){}\put(16,1){{\tiny $ 2$}}
\multiframe(0,7.5)(13.5,0){1}(35,7){}\put(38,9){\tiny{$ s-1$}}
\end{picture}\,\,
\oplus
\begin{picture}(55,15)(-5,10)
\multiframe(0,0)(13.5,0){1}(7,7){}\put(9,1){{\tiny $ 1$}}
\multiframe(0,7.5)(13.5,0){1}(7,7){}\put(9,8){\tiny{$ 1$}}
\multiframe(0,15)(13.5,0){1}(35,7){}\put(38,15){\tiny{$ s-1$}}
\end{picture}\,\,
\oplus
\begin{picture}(43,15)(-5,5)
\multiframe(0,0)(13.5,0){1}(7,7){}\put(9,1){{\tiny $ 1$}}
\multiframe(0,7.5)(13.5,0){1}(35,7){}\put(38,9){\tiny{$ s$}}
\end{picture}\,\,
/
\begin{picture}(55,15)(-5,5)
\multiframe(0,0)(13.5,0){1}(7,7){}\put(9,1){{\tiny $ 1$}}
\multiframe(0,7.5)(13.5,0){1}(35,7){}\put(38,9){\tiny{$ s-2$}}
\end{picture}\,\,
\oplus
\begin{picture}(50,12)(0,0)
\multiframe(0,0)(13.5,0){1}(30,7){}\put(33,1){\tiny{$ s-1$}}

\end{picture}.
\label{oY3}\ee
where $/$ means that the last two diagrams, which take into account
the relaxed traceless condition (\ref{trnn}), must be subtracted
from the first three.

The \emph{traceful} Young tableau decomposition of the gauge
parameter $\xi_{a_1\cdots a_{s-1},bm}$ satisfying eq. (\ref{trmn})
is
\be
\begin{picture}(09,13)
{
\put(05,05){\line(1,0){05}}%
\put(05,10){\line(1,0){05}}%
\put(05,0){\line(1,0){05}}%
\put(05,0){\line(0,1){10}}%
\put(10,0.0){\line(0,1){10}}
}
\end{picture}\,\,
\oplus
\begin{picture}(55,15)(-5,5)
\multiframe(0,0)(13.5,0){1}(14,7){}\put(16,1){{\tiny $ 2$}}
\multiframe(0,7.5)(13.5,0){1}(35,7){}\put(38,9){\tiny{$ s-1$}}
\end{picture}\,\,
/
\begin{picture}(55,15)(-5,5)
\multiframe(0,0)(13.5,0){1}(7,7){}\put(9,1){{\tiny $ 1$}}
\multiframe(0,7.5)(13.5,0){1}(35,7){}\put(38,9){\tiny{$ s-2$}}
\end{picture}\,\,
\oplus
\begin{picture}(50,12)(0,0)
\multiframe(0,0)(13.5,0){1}(30,7){}\put(33,1){\tiny{$ s-1$}}
\end{picture}\,.
\label{xi3}\ee
It is obtained by subtracting from the traceful Young tableau
\be
\begin{picture}(55,15)(-5,5)
\multiframe(0,0)(13.5,0){1}(14,7){}\put(16,1){{\tiny $ 2$}}
\multiframe(0,7.5)(13.5,0){1}(35,7){}\put(38,9){\tiny{$ s-1$}}
\end{picture}\,\,
\ee
the Young tableaux corresponding to the relaxed trace condition
(\ref{trmn})
\be\label{stolbik}
\begin{picture}(55,15)(-5,5)
\multiframe(0,0)(13.5,0){1}(7,7){}\put(9,1){{\tiny $ 1$}}
\multiframe(0,7.5)(13.5,0){1}(35,7){}\put(38,9){\tiny{$ s-2$}}
\end{picture}\,\,
\oplus
\begin{picture}(50,12)(0,0)
\multiframe(0,0)(13.5,0){1}(30,7){}\put(33,1){\tiny{$ s-1$}}
\end{picture}\,
/
\begin{picture}(09,13)
{
\put(05,05){\line(1,0){05}}%
\put(05,10){\line(1,0){05}}%
\put(05,0){\line(1,0){05}}%
\put(05,0){\line(0,1){10}}%
\put(10,0.0){\line(0,1){10}}
}
\end{picture}\,\,.
\ee
In (\ref{stolbik}), in the case of the odd spins, the Young tableau
\begin{picture}(09,13)
{
\put(05,05){\line(1,0){05}}%
\put(05,10){\line(1,0){05}}%
\put(05,0){\line(1,0){05}}%
\put(05,0){\line(0,1){10}}%
\put(10,0.0){\line(0,1){10}}
}
\end{picture}\,,
is subtracted, since it is not part of the traceless condition
\be\label{stolbik1}
\eta^{a_{s-1}b}\,\xi_{a_1\cdots a_{s-1},bm}=0\,,
\ee
which one can see by considering the symmetry of the full trace of
(\ref{stolbik1}) in the indices $a_1\cdots a_{s-3}$.

 Comparing (\ref{xi3}) with (\ref{oY3}) we conclude that the
number of the components of the connection which are not gauged away
by the Stueckelberg symmetry is the same as the number of the
torsion components modulo an antisymmetric tensor field $F_{mn}$
corresponding to
\begin{picture}(09,13)
{
\put(05,05){\line(1,0){05}}%
\put(05,10){\line(1,0){05}}%
\put(05,0){\line(1,0){05}}%
\put(05,0){\line(0,1){10}}%
\put(10,0.0){\line(0,1){10}}
}
\end{picture}\,,
which is thus a pure gauge. This makes one more evidence to the fact
stressed in the main text that our construction does not include the
massless field of spin 1 whose field strength would be $F_{mn}$.

\def\thesection{}
\def\theequation{B.\arabic{equation}}
\section{Appendix B}
\setcounter{equation}0

As an alternative to the approach explained in the main text, let us
explain how to find the general solution of the equation
(\ref{meq}). In this approach, however, the problem that remains to
be solved is to find appropriate analytic functions $A(\s,\t,v)$
such that the resulting solution $\tilde{\Phi}(\s,\t,v)$ be free of
the constant parts in $\s$ and $\t$.

The generic solution $\tilde{\psi}_0(\s,\t,v)$ of the homogeneous
first-order partial differential equation (\ref{meq}) with
$\tilde{A}(\s,\t,v)=0$ is
 \be \label{finc}
\tilde{\psi}_0(\s,\t,v)= \t \Big
(1-\frac{2\t}{\s}\Big )^{-\frac{D-5}{2}} (1-\t^4
v)^{\frac{D-9}{2}}\tilde{\psi}\Big (\xi, v\Big)\,,
\ee
where
\be
\label{xi} \xi = \frac{1-\t^4 v}{\s\t(1-\frac{2\t}{\s})} -\t^2v
\ee
and $\tilde{\psi}(\xi,v)$ is an arbitrary function of its arguments.

For the first sight it may look as we have constructed a topological
action with the trivial variation. This is not the case, however. As
expected, the action that contains $\Phi$ defined by (\ref{finc})
via (\ref{rel11})-(\ref{rel31}) is identically zero because the
expansion of the function $\Phi$  in power series
\be\label{Phi} \Phi(u,w,v)=\oint d\s d\t e^{\s u+\t
w} \sum_{p,q,r}\t^p \s^q  v^r
\ee
contains only the terms with $p>q$,\emph{ i.e.} those with more $\s$
than $\t $ in the denominator\footnote{Recall that; because of the
definition of the measure (\ref{res}) only the terms with negative
$p$ and $q$ contribute to eq. (\ref{Phi}).}. All such terms do not
contribute because of the Young property of fields. Indeed, they
describe terms in which more than half of the indices of a tensor
are contracted with the compensator contained in $u$ (\ref{u}). This
implies the symmetrization over more than half of indices of the
tensor described by a rectangular Young tableau, thus giving zero.
Note that the dependence on $ v$ does not affect this argument
because its application does not affect the Young symmetry property,
mapping two-row Young tableaux to shorter two-row Young tableaux.

Thus, as expected, the  solution of the homogeneous equation
describes nothing. Setting
\be \tilde{\psi}(\s,\t,v)= \frac{\t (\s\t)^{ \frac{D-5}{2}}
(1-\t^4 v)^{ \frac{D-9}{2}}}{(\s\t -2\t^2)^{
\frac{D-5}{2}}}\chi(\xi,\t,v)\,,
\ee
changing the variables from $\s$, $\t$ to $\xi$ (\ref{xi}) and $y$
\be y=2+\frac{1}{t^2\xi}\,,
\ee
 where $v$ should be interpreted as a parameter, and using the
relations
\be \s\t -2\t^2  = \frac{1- \frac{v}{(2-y)^2\xi^2}}
{\xi(1- \frac{v}{(2-y)\xi^2})}\q \s\t= \frac{-\frac{y}{2-y}
+\frac{v}{\xi^2 (2-y)^2}}{\xi(1- \frac{v}{(2-y)\xi^2})} \,,
\ee
the inhomogeneous equation amounts to
\be \label{eqchi}
-(2-y)\frac{\partial }{\partial y} \chi(\xi,\t,v) = \frac{\xi
(1-\frac{v}{(2-y)^2\xi^2})(1-\frac{v}{(2-y)\xi^2})} {(-\frac{y}{2-y}
+\frac{v}{\xi^2 (2-y)^2})^{\frac{D-3}{2}}} A(y,\xi,v)\,.
\ee
The appropriate solution of this equation that admits an expansion
in integer negative powers of $\s$ and $\t$  is
\be
\chi(\s,\t,v)=\int^1_0 \frac{dt}{t} \frac{y}{2t  -y}\,
\frac{\xi
(1-\frac{v}{(2-t^{-1}y)^2\xi^2})(1-\frac{v}{(2-t^{-1}y)\xi^2})}
{(-\frac{y}{2t-y} +\frac{v}{(2-t^{-1}y)^2\xi^2})^{\frac{D-3}{2}}}
A(yt^{-1}, \xi, v)\,.
\ee
With the help of the relations (\ref{rel11})-(\ref{rel31}) we obtain
that
\be \tilde{\Phi}(\s,\t,v)=-2\frac{(1-\t^4v)^{\frac{D-7}{2}}}
{\xi (2-y) (1-2\frac{\t}{\s})^{\frac{D-5}{2}}}
\chi(\s,\t,v)\,. \ee It is also easy to
reconstruct the functions $\Lambda(\s,\t,v)$ and $W(\s,\t,v)$.

\end{document}